\documentclass[10pt,a4paper]{article}

\usepackage{caption}
\usepackage[font=footnotesize,labelformat=simple]{subcaption}
\usepackage{amssymb,amsmath,graphicx,color,amsthm} 

\usepackage[authoryear]{natbib}
\usepackage{dcolumn} 

\usepackage{float} 
\usepackage{bbm}
\usepackage[colorlinks=true,linkcolor=blue,citecolor=red]{hyperref}
\usepackage{bbding}
\usepackage{pifont}
\usepackage{wasysym}
\usepackage{bbding}
\usepackage{amsfonts}
\usepackage[ruled,vlined]{algorithm2e}

\usepackage{bm}

\usepackage{lscape}
\usepackage{pgf}
\usepackage{verbatim}
\usepackage{multirow}
\usepackage{enumerate}
\usepackage{comment}
\usepackage[utf8]{inputenc} 
\usepackage[T1]{fontenc}    
\usepackage{hyphenat}

\usepackage[a4paper]{geometry}
\title{\textbf{Imputation of Missing Data Using Linear Gaussian Cluster-Weighted Modeling}}
\author{Luis Alejandro Masmela-Caita\footnote{Universidad Distrital Francisco José de Caldas. Bogotá - Colombia. e-mail: lmasmela@udistrital.edu.co}\\
	Thais Paiva Galletti\footnote{Universidade Federal de Minas Gerais. Belo Horizonte - MG - Brasil. e-mail: thaispaiva@est.ufmg.br}\\
	Marcos Oliveira Prates\footnote{Universidade Federal de Minas Gerais. Belo Horizonte - MG - Brasil. e-mail: marcosop@est.ufmg.br}
}

\date{}

\begin{document}

\maketitle

\begin{abstract}
Missing data theory deals with the statistical methods in the occurrence of missing data. Missing data occurs when some values are not stored or observed for variables of interest. However, most of the statistical theory assumes that data is fully observed. An alternative to deal with incomplete databases is to fill in the spaces corresponding to the missing information based on some criteria, this technique is called imputation. We introduce a new imputation methodology for databases with univariate missing patterns based on additional information from fully-observed auxiliary variables. We assume that the non-observed variable is continuous, and that auxiliary variables assist to improve the imputation capacity of the model. In a fully Bayesian framework, our method uses a flexible mixture of multivariate normal distributions to model the response and the auxiliary variables jointly. Under this framework, we use the properties of Gaussian Cluster-Weighted modeling to construct a predictive model to impute the missing values using the information from the covariates. Simulations studies and a real data illustration are presented to show the method imputation capacity under a variety of scenarios and in comparison to other literature methods.\\


\end{abstract}

\section{Introduction}
In studies with statistical data, the vast majority of inference methods start with the premise that the database has its information completely observed. Moreover, usually, the properties of the estimators relies on this assumption.
However, in practice, databases with complete information cannot be guaranteed due to multiple reasons, for example, when information is collected through surveys. These reasons must be studied in depth to determine why data is missing or how to avoid incomplete data sets in the information gathering process. To perform statistical analysis with complete databases, the most common solution is to remove individuals with missing information from the database (\textit{Procedures Based on Completely Recorded Units}). Besides the information loss, this approach can lead to estimation biases, particularly when there are differences between the information of those who respond and those who do not. \citet{rubin1976inference} discusses the conditions of when the process that caused the missing observations can be ignored. He presents the weakest conditions in the missing data process so that it is appropriate to ignore this process when making inferences about the data distribution.\\

An alternative is to deal with incomplete databases by filling in the spaces corresponding to this missing information based on some criteria, this technique is called \textit{imputation}. 
Imputation is attractive because it allows the use of methods designed for complete data sets. 
The \textit{Expectation-Maximization} (EM) algorithm \citep{dempster1977maximum} 
made it possible to generate robust estimators from the application of the \textit{Maximum Likelihood} (ML) method, where the missing observations are assumed as random variables and the imputed data are generated without the need to fit 
models. One drawback of imputation, followed by the use of full data set analysis methods, is that the resulting inferences can be misleading if the uncertainty due to lack of data has not been properly addressed \citep{little2019statistical}. In this regard, \citet{rubin1987multiple} introduces the general approach of \textit{Multiple Imputation} (MI), based on the premise that each missing data must be replaced by various simulated values. 
Multiple imputation is motivated by the Bayesian framework and as such, the general methodology suggested for imputation is to impute using the posterior predictive distribution of the missing data given the observed data and some estimate of the parameters \citep{jamshidian2007advances}.\\

A complete study of the imputation techniques and their classification can be reviewed in \citet{little2019statistical}. Imputation techniques are also used to generate synthetic data sets in the case of data with disclosure restrictions. Since the synthetic values are not actual observations, they can be published for analysis \citep{rubin1993statistical,raghunathan2003multiple}. The use of finite mixture models provides a flexible approach for statistical modeling of a wide variety of random phenomena. In the case of incomplete data, for example, \citet{ghahramani1995learning} establish an algorithm for the estimation of the components of the mixture. 
\citet{di2007imputation} propose using a finite mixture of multivariate Gaussian distributions to handle imputation of missing data. 
By using the Expectation-Maximization (EM) algorithm, the model is estimated considering the missing information, and then, for the imputation process, two alternatives are used, the conditional mean imputation and the random selection. \citet{liao2007quadratically} present the Quadratically Gated Mixture of Experts in classification of incomplete data. The EM algorithm is used for joint probability maximization, with adaptive imputation performed analytically in step E. \citet{sovilj2016extreme} explore the general regression framework for missing data. To provide reliable estimates for the regression function, a methodology is developed in which the Gaussian mixture is used to model the data distribution when values are missing. At the same time, Extreme Learning Machine is used to handle multiple imputation.\\

Of particular interest, for information obtained through surveys, \citet{kim2014multiple} propose a fully Bayesian, flexible joint modeling approach for multiple imputation of missing or faulty data subject to linear constraints. The procedure is based on a Dirichlet process mixture of multivariate normal distributions as the base imputation engine. The missing data are imputed with values generated from the fitted model to the observed data. Using a similar statistical model as an imputation engine, \citet{paiva2017stop} propose a methodology to impute continuous variables with missing data, where the missing data mechanism is non-ignorable. Under a Bayesian approach, the procedure begins by fitting a mixture of multivariate normal distributions based on the observed data. Then, from subsequent samples of the mixture model, an analyst can use the fitted model 
to obtain imputed data in various scenarios.\\


Our interest in this work lies in developing a flexible imputation model with auxiliary information from fully observed variables. To do so, we follow
the context of mixture models \citep{fruhwirth2006finite, mclachlan2004finite}. The implementation of mixture regression models was initially studied in the context that 
fully observed variables assume the role of covariates. The covariates are treated as deterministic, so they do not carry information about which mixture component the subject is likely to belong to. Although this assumption may be reasonable in experiments where the explanatory variable is completely determined by the experimenter, \citet{hoshikawa2013mixture} states that with observational data the covariates may behave differently between groups. Therefore, the model must also incorporate the heterogeneity of the covariates, which allows estimating the subject component from this information. A model that has 
these characteristics is the \textit{Cluster-Weighted Modeling} (CWM), developed by \citet{gershenfeld1997nonlinear} in the context of media technology. \citet{ingrassia2012local} proposed to use the CWM in a statistical environment and showed that it is a general and flexible family of mixture models. In particular, under Gaussian assumptions, they specify characteristics of their probability distribution and related statistical properties and demonstrate links with traditional mixture models in terms of density functions and posterior probabilities. An interesting result is that the Gaussian CWM includes the \textit{Finite Mixture Model} (FMM) and the \textit{Mixture of Regression Model} (MRM) as special cases.\\

\begin{figure}[h]
	\centering
	\subcaptionbox{No auxiliary information.
		\label{base_1}}
	{\includegraphics[scale=0.250]{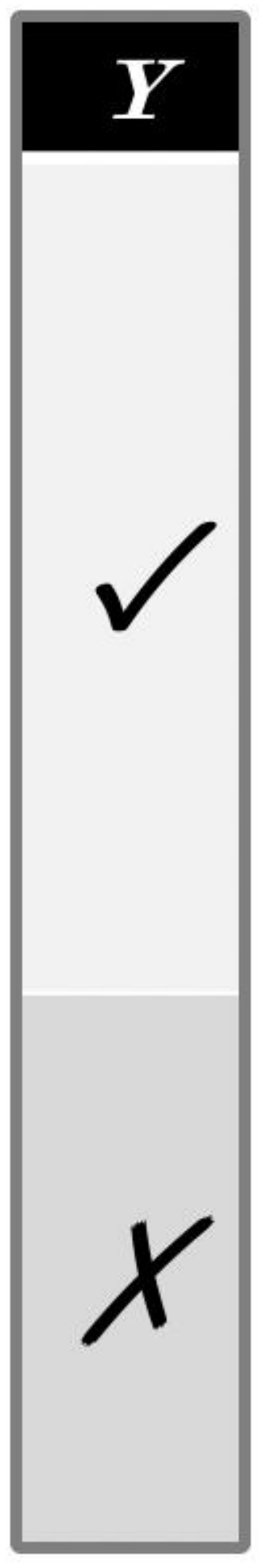}}
	\subcaptionbox{With auxiliary information.
		\label{base_2}}
	{\includegraphics[scale=0.250]{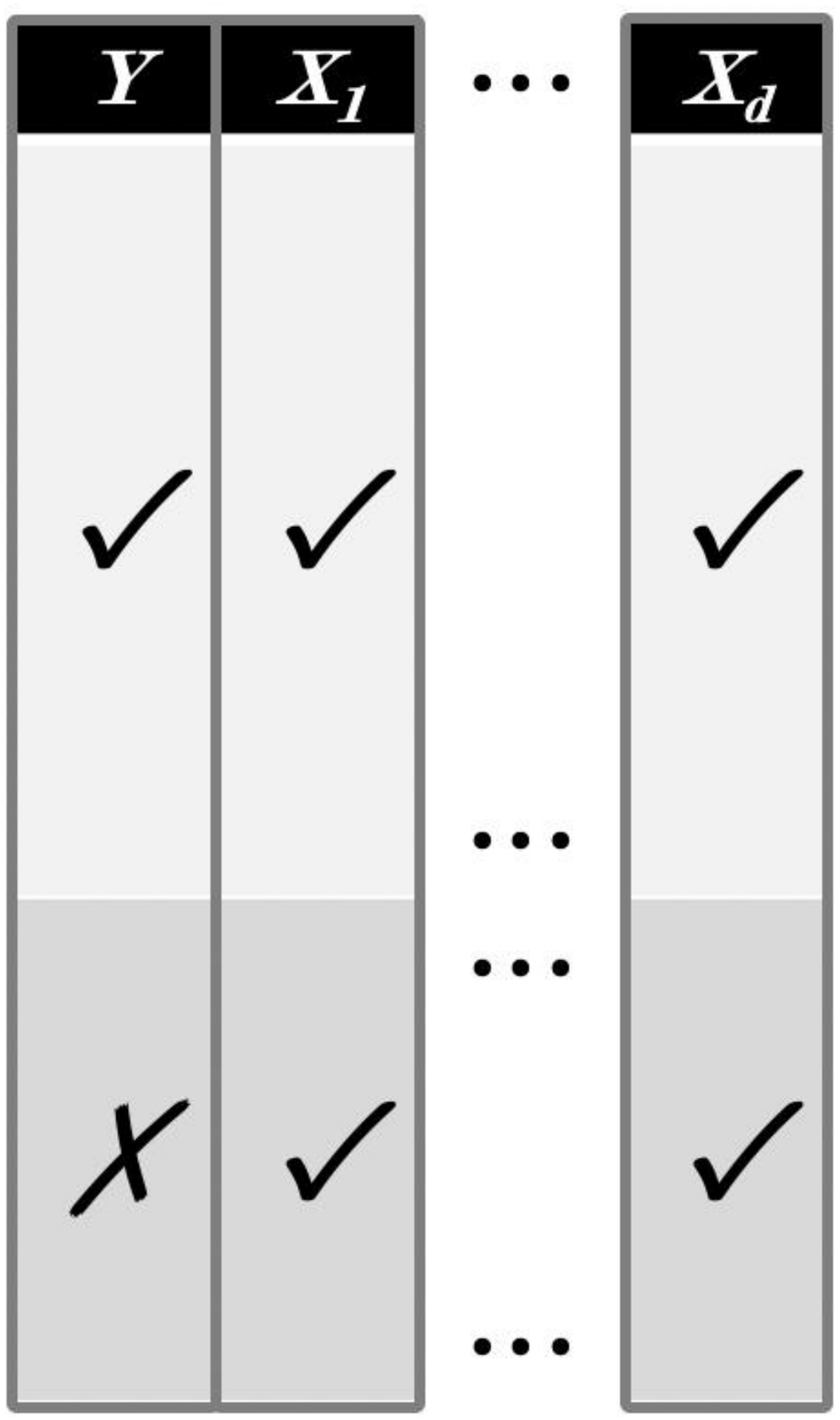}}
	\caption[]{Missing data pattern from univariate databases in the cases of not including and including auxiliary information.}
	\label{bases}
\end{figure}

Our objective is to propose a new imputation procedure that includes information from other sources. 
Following \citet{gershenfeld1999nature}, in the CWM context we call the output variable the variable with missing information, while those that are observed entirely and that we want to include as new information will be called input variables. In Figure \ref{bases}, we illustrate the missing data patterns from the databases we want to deal with. In them, the check mark ({\footnotesize \Checkmark}) indicates that the data is observed, while the tag ({\footnotesize \XSolidBrush}) indicates that the data is missing. Figure \ref{base_1} refers to the pattern of missing data from which we started and that does not consider auxiliary information. A mixture of multivariate normal distributions combined with a Dirichlet process to determine the number of components is implemented as an imputation engine \citep{kim2014multiple, paiva2017stop}. Assuming that additional information can be obtained from auxiliary data sources, the missing data pattern can take the form shown in Figure \ref{base_2}. In this new pattern, it is observed that new variables $X_1,...,X_d$ appear with fully observed information for all individuals. We propose to use the extra information available so that the imputations for individuals with missing values in the response variable $Y$ improve under some criteria. The performance evaluation of the model will be carried out taking into account the type of information provided by the input variables, and comparing it to the performance of other imputation methods of interest.\\

The remainder of this article is distributed as follows. In Section~\ref{sec:GLCWM}, we present the Cluster-Weighted Modeling specified for the Gaussian case. Some results are presented relating the different mixture models of interest. In Section \ref{sec:metodologia}, the estimation and imputation procedures of the model are presented from a totally Bayesian perspective. An algorithm that wraps both procedures is presented and summarizes the basis of our model. Next, Section \ref{sec:estudios_simulacion} shows the simulation studies established for the purpose of analyzing our model. The first part aims to evaluate the performance of the LCWM against the different types of input variables. The second part of the simulation procedure compares the performance of our model with other methods used in the imputation process. These are predictive methods based on Bayesian approaches. In Section~\ref{sec:ejemplo}, the Faithful database is used to illustrate the performance of the proposed imputation methodology in a real data. Finally, Section~\ref{sec:conclusiones} concludes with a discussion and suggestions for future work.\\

\section{Preliminaries}
\label{sec:GLCWM}

\textit{Finite Mixture Models} (FMM) are used to treat heterogeneous data in various experimental situations, for example, when measurements of the random variable are taken in two or more different conditions. This can be interpreted as if the information came from sub-populations that are called components. Obtaining these components leads to the estimation of the parameters of the mixture. Several textbooks and documents on studies of the theory of FMM can be consulted \citep{fruhwirth2006finite,mclachlan2004finite,nguyen2015finite}.\\

\textit{Cluster-Weighted Modeling} (CWM) is a flexible procedure that seeks to model the joint probability of data that comes from heterogeneous populations. The interest is the pair $(\bm{X}',Y)'$ constituted by the random vector $\bm{X}\in \mathbb{R}^d$ and the random variable $Y\in \mathbb{R}$. CWM was initially introduced by \citet{gershenfeld1997nonlinear} for modeling time series data related to musical instrument parameters. Within the framework of the Gaussian \textit{Mixtures of Regressions Models} (MRM), \citet{ingrassia2012local} proposes its use in a statistical environment and showed that it is a general and flexible family of mixture models.\\

The CWM decomposes the joint probability $p(\bm{x},y)$ as follows,
\begin{equation}
p(\bm{x},y)=\sum_{Z=1}^{G} p(y|\bm{x},Z) \, p(\bm{x}|Z) \, \alpha_Z,
\label{cwm}
\end{equation}
where $p(y|\bm{x},Z)$ is the conditional density of the response variable $Y$ given the predictor vector $\bm{X}$ in the component indicated by $Z$, $p(\bm{x}|Z)$ is the probability density of the variable $\bm{X}$ in the group $Z$, and $\alpha_Z$ is the sampling probability of the group marked with $Z$. Therefore, the joint density of $(\bm{X}',Y)'$ can be seen as a mixture of local models $p(y| \bm{x},Z)$ weighted (in a broader sense) by both local densities $p(\bm{x}|Z)$, and the mixing weights $\alpha_Z$.\\

For applications whose purposes are classification, the interest is focused on the posterior probability $p(Z|\bm{x},y)$ that the observation $(\bm{x}',y)'$ belongs to the component $Z$ given by,
\begin{equation}
	p(Z|\bm{x},y)=\frac{\alpha_Zp(\bm{x}|Z)p(y|\bm{x},Z)}{\sum_{Z=1}^{G}\alpha_Zp(\bm{x}|Z)p(y|\bm{x},Z)},
	\label{pesos_CWM}
\end{equation}
that is, the classification of each observation depends on both the marginal and conditional densities. Furthermore, 
it can also be obtained that,
\begin{equation*}
p(Z|\bm{x},y)=\frac{p(Z|\bm{x})p(y|\bm{x},Z)}{\sum_{Z=1}^{G}p(Z|\bm{x})p(y|\bm{x},Z)},
\end{equation*}
with,
\begin{equation}
p(Z|\bm{x})=\frac{\alpha_Zp(\bm{x}|Z)}{\sum_{Z=1}^{G}\alpha_Zp(\bm{x}|Z)},
\label{pesos_mezcla_x}
\end{equation}
where $p(Z|\bm{x})$ is the posterior probability that the observation $\bm{x}$ belongs to the component $Z$.\\

If in expression \eqref{cwm} we call $\bm{w} = (\bm{x}', y)'$ and we assume within each component that $p(\bm{w}|Z)=p((\bm{x}', y)'|Z)= p(y|\bm{x},Z) \, p(\bm{x}|Z)$ then this density takes the form,
\begin{equation*}
p(\bm{w})=\sum_{Z=1}^{G} p(\bm{w}|Z) \, \alpha_Z,
\end{equation*}
which refers to FMM.

\subsection{Gaussian Linear CWM}
The 
Gaussian \textit{Linear Cluster-Weighted Modeling} (LCWM) is based on considering the marginal and conditional distributions as normal distributions. That is, $p(\bm{x}|Z)=\phi_d(\bm{x};\bm{\mu}_Z,\Sigma_Z)$, while $p(y|\bm{x},Z)=\phi_1(y;\mu(\bm{x},\bm{\beta}_Z),\sigma^2_Z)$, where the conditional density is based on linear mappings, i.e. $\mu(\bm{x},\bm{\beta}_Z)=\bm{b}'_Z\bm{x}+b_{Z,0}$, for some $\bm{\beta}_Z=(\bm{b}'_Z,b_{Z,0})'$, with $\bm{b}_Z \in \mathbb{R}^d$ and $b_{Z,0} \in \mathbb{R}$. Under these conditions, the expression in \eqref{cwm} takes the following form
\begin{equation}
p(\bm{x},y)=\sum_{Z=1}^{G}\phi_d(\bm{x};\bm{\mu}_Z,\Sigma_Z) \, \phi_1(y;\bm{b}'_Z\bm{x}+b_{Z,0},\sigma^2_Z) \, \alpha_Z.
\label{lcwmg}
\end{equation}
The density for the Gaussian LCWM is given by the expression in \eqref{lcwmg}. In this case, the posterior probability in \eqref{pesos_CWM} can be written as,
\begin{equation}
p(Z|\bm{x},y)=\frac{\phi_d(\bm{x};\bm{\mu}_Z,\Sigma_Z)\phi_1(y;\bm{b}'_Z\bm{x}+b_{Z,0},\sigma^2_Z)\alpha_Z}{\sum_{Z=1}^{G}\phi_d(\bm{x};\bm{\mu}_Z,\Sigma_Z)\phi_1(y;\bm{b}'_Z\bm{x}+b_{Z,0},\sigma^2_Z)\alpha_Z},
\label{pesos_LCWMG}
\end{equation}
while, the expression in \eqref{pesos_mezcla_x} takes the form,
\begin{equation}
p(Z|\bm{x})=\frac{\phi_d(\bm{x};\bm{\mu}_Z,\Sigma_Z)\alpha_Z}{\sum_{Z=1}^{G}\phi_d(\bm{x};\bm{\mu}_Z,\Sigma_Z)\alpha_Z}.
\label{pesos_LCWMG_x}
\end{equation} 

\subsection{Some relationships between FMM, MRM, and LCWM in the Gaussian case}
\label{sec:rel_models}


In this section, three results about the probability functions and the posterior probabilities are presented to relate the FMM, MRM and LCWM in the Gaussian case. The propositions and their respective proofs can be consulted in \citet{ingrassia2012local,nguyen2015finite}.\\

Let $\bm{W}$ be a random vector that takes values in $\mathbb{R}^{d+1}$ with joint probability distribution $p(\bm{w})$. Assume that the density $p(\bm{w})$ corresponds to an FMM, that is,
\begin{equation*}
p(\bm{w})=\sum_{Z=1}^{G} p(\bm{w}|Z) \, \alpha_Z,
\end{equation*}
where $p(\bm{w}|Z)$ is the probability density of $\bm{W}$ given $Z$, and $\alpha_Z=p(Z)$ is the mixture weight of the components marked with $Z \in \{1,. . . , G\}$. In the case of a Gaussian FMM, let $\bm{\mu}^{(\bm{w})}_Z$ and $\Sigma^{(\bm{w})}_Z$ be the vector of means and the covariance matrix of $\bm{W}|Z$, respectively.
Suppose that $\bm{W}=(\bm{X}',Y)'$, where $\bm{X}$ is a random vector taking values in $\mathbb{R}^{d}$ and $Y$ a random variable. Then,
\begin{equation*}
\bm{\mu}_Z^{(\bm{w})}=
\begin{pmatrix}
\bm{\mu}_Z^{(\bm{x})}\\
\mu_Z^{(y)}
\end{pmatrix}
\quad \text{ and } \quad \Sigma_Z^{(\bm{w})}=
\begin{pmatrix}
\Sigma_Z^{(\bm{xx})} & \Sigma_Z^{(\bm{x}y)} \\
\Sigma_Z^{(y\bm{x})} & \sigma_Z^{2(y)} 
\end{pmatrix}.
\label{parametros_FMG}
\end{equation*}
A first interesting result indicates that the CWM contains the FMM and, specifically, in the Gaussian context, restricting the CWM to the case LCWM, FMM and LCWM are equivalent \citep{ingrassia2012local}.\\

From the proof of this result it is worth noting how the density, 
	\begin{equation*}
	p(\bm{w})=\sum_{Z=1}^{G} \phi_{d+1}\left(\bm{w};\bm{\mu}_Z^{(\bm{w})},\Sigma_Z^{(\bm{w})}\right)\alpha_Z,
	\end{equation*}
is written to bring it into the structure of a Gaussian LCWM. So,
\begin{equation*}
p(\bm{w})=\sum_{Z=1}^{G}\phi_d\left(\bm{x};\bm{\mu}_Z^{(\bm{x})},\Sigma_Z^{(\bm{xx})}\right)\mathcal\phi_1\left(y|\bm{x};\mu_Z^{(y|\bm{x})},\sigma_Z^{2(y|\bm{x})}\right)\alpha_Z,	
\end{equation*}
where,
\begin{equation*}
\begin{split}
\mu^{(y|\bm{x})}_Z&=\mu^{(y)}_{Z}+\Sigma^{(y\bm{x})}_{Z} {\Sigma^{(\bm{xx})}_{Z}}^{-1}\left(\bm{x}-\bm{\mu}^{(\bm{x})}_{Z}\right)\\
&=\left[\mu^{(y)}_{Z}-\Sigma^{(y\bm{x})}_{Z} {\Sigma^{(\bm{xx})}_{Z}}^{-1}\bm{\mu}^{(\bm{x})}_{Z}\right]+\left[\Sigma^{(y\bm{x})}_{Z} {\Sigma^{(\bm{xx})}_{Z}}^{-1}\right]\bm{x}\\
&=b_{Z,0}+\bm{b}_{Z}\bm{x},
\end{split}
\end{equation*}
and,
\begin{equation*}
\begin{split}
\sigma_Z^{2(y|\bm{x})}&=\sigma_Z^{2(y)}-\Sigma^{(y\bm{x})}_{Z} {\Sigma^{(\bm{xx})}_{Z}}^{-1}\Sigma_Z^{(\bm{x}y)}\\
&=\sigma^2_Z.
\end{split}
\end{equation*} 
It can also be shown that FMM and LCWM have the same posterior probability distribution.\\

The second result involves the MRM in the Gaussian case given by,
\begin{equation}
	p(y|\bm{x})=\sum_{Z=1}^{G}\phi_1(y;\bm{b}'_Z\bm{x}+b_{Z,0},\sigma^2_Z) \alpha_Z,
	\label{GLMRM}
\end{equation}
and for which, the posterior probability $p(Z|\bm{x},y)$ is given by the expression,
\begin{equation}
p(Z|\bm{x},y)=\frac{\phi_1(y;\bm{b}'_Z\bm{x}+b_{Z,0},\sigma^2_Z)\alpha_Z}{\sum_{Z=1}^{G}\phi_1(y;\bm{b}'_Z\bm{x}+b_{Z,0},\sigma^2_Z)\alpha_Z}.
\label{pesos_LMRM}
\end{equation}  

This result establishes an expression that relates the LCWM and the MRM in the Gaussian context, when the covariate $\bm{x}$ has the same behavior among components. Specifically, when $\phi_d(\bm{x};\bm{\mu}_Z,\Sigma_Z)=\phi_d(\bm{x};\bm{\mu},\Sigma)$ for all $Z \in \{1,. . . , G\}$, then it follows that,
\begin{equation}
	p(\bm{x},y)=\phi_d(\bm{\mu},\Sigma)p(y|\bm{x}),
\end{equation}
where $p(y|\bm{x})$ is the Gaussian MRM in \eqref{GLMRM}.\\

Finally, under the assumption that the covariate $\bm{x}$ has the same behavior between components, a third result establishes the equality between the posterior probabilities for LCWM and MRM in the Gaussian case, i.e., the expression in \eqref{pesos_LCWMG} coincides with \eqref{pesos_LMRM}. Additionally, under the same conditions 
we have that the posterior probabilities in \eqref{pesos_LCWMG_x} simplify to $p(Z|\bm{x})=\alpha_Z$.\\

\section{Methodology}
\label{sec:metodologia}
\subsection{Bayesian estimation of the model} 
\label{subsec:Bayesian_est}

The first result in Section~\ref{sec:rel_models} establishes a mapping between the parameter vectors of the FMM and the LCWM in the Gaussian context, therefore, the interest initially lies in the estimation of the FMM parameters. A  Bayesian approach is implemented on a mixtures of multivariate normal distributions using the stick-breaking representation of a truncated Dirichlet process \citep{ferguson1973bayesian, sethuraman1994constructive} as the prior distribution of the mixing weights.\\ 



Suppose that each individual in the data set belongs to one of $G$ mixture components, that is, $Z_i \in \{1,...,G \}$ so that $\bm{Z}=(Z_1,...,Z_n)$. The mixing probabilities are given by $\bm{\alpha}=(\alpha_1,...,\alpha_G)$ with $\alpha_g = P(Z_i=g)$ where $i=1,...,n$ and $g=1,...,G$.\\
If $\bm \mu= (\bm \mu_1,...,\bm \mu_G)$ and $\bm\Sigma=(\Sigma_1,...,\Sigma_G)$ then,
\begin{equation*}
\begin{split}
\textbf{\textit{w}}_i|Z_i,\bm\mu,\bm\Sigma &\sim \mathcal{N}_p(\bm \mu_{Z_i},\Sigma_{Z_i}),\\
Z_i|\bm \alpha &\sim \text{Multinomial}(\bm \alpha). 
\end{split}
\end{equation*}


For the vector of mixture parameters $\bm\alpha$, we adopted as prior distribution a stick-breaking representation of a truncated Dirichlet process. It has been shown that this representation was successful in other mixture model applications for multiple imputation of missing data \citep{si2013nonparametric,manrique2017bayesian}, and it allows for fast MCMC convergence \citep{kim2014multiple}. So, using this representation, we have that
\begin{equation*}
\begin{split}
\alpha_g &= \nu_g \prod_{k<g} (1-\nu_k) \text{ for } g=1,...,G,\\
\nu_g &\sim \text{Beta}(1,\eta) \text{ for } g=1,...,G-1; \quad \nu_G=1,\\
\eta &\sim \text{Gamma}(a_{\eta},b_{\eta}),
\end{split}
\end{equation*}
and let the prior specification of $(\bm\mu,\bm\Sigma)$ be 
\begin{equation*}
\begin{split}
\bm{\mu}_g|\Sigma_g &\sim \mathcal{N}_p(\bm \mu_0,h^{-1}\Sigma_g),\\
\Sigma_g &\sim \text{Inverse Wishart}(f,\Delta),
\end{split}
\end{equation*}
where $f$ is the prior degrees of freedom and $\Delta=\text{diag}(\delta_1,...,\delta_p)$ with $\delta_j \sim \text{Gamma}(a_{\delta},b_{\delta})$ for $j=1,...,p.$\\ 


It is possible to use a Gibbs sampler algorithm to estimate the posterior distribution. For the number of components $G$, \citet{kim2015simultaneous} recommend to set it to a somewhat large value, for example $G=30$.
At each iteration of the Gibbs sampler, the number of nonempty components is counted. If this count reaches the value assigned to $G$, it is prudent to increase $G$ and readjust the model with more components. When the count of nonempty components is less than $G$, then the selected value of $G$ is reasonable.

\subsection{Imputation procedure}

Since our objective is to establish an imputation procedure that starts from the structure of the missing data pattern in Figure~\ref{bases} and that includes auxiliary information observed for all individuals, we will now describe the development of this procedure.\\

Assuming observational data, we use the Bayesian approach presented in Section~\ref{subsec:Bayesian_est} to fit a Gaussian FMM using joint information from the input and output variables. Next, based on the results of Section~\ref{sec:GLCWM}, we obtain the estimates for the Gaussian LCWM parameters through which we establish the imputation step. Using the values of the input variables corresponding to the individuals with missing information, it is possible to calculate posterior probabilities and use them adaptively to decide with which of the components to impute.\\

The algorithm starts by assigning initial values for the missing data in the output variable, generating a complete database. Then, values of $\bm{Z}$ are drawn to assign each observation to a component, and thus, obtain initial estimates of the parameters $\bm{\alpha}$, $\bm{\mu}$, and $\bm{\Sigma}$ that characterize the Gaussian FMM. The first result presented in Section~\ref{sec:rel_models} allows obtaining, based on the parameters $\bm{\alpha}$, $\bm{\mu}$, and $\bm{\Sigma}$, values of the parameters corresponding to the Gaussian LCWM. More specifically, initial estimates of the parameters are obtained for the marginal and conditional distributions involved in the expression \eqref{lcwmg}, as well as for the posterior probabilities in \eqref{pesos_LCWMG} and \eqref{pesos_LCWMG_x}. When implementing the Gibbs sampler algorithm, posterior probabilities are of special interest. The one specified by expression \eqref{pesos_LCWMG}, allows, within each iteration, to obtain values of $\bm{Z}$ for complete observations, and estimates of the parameters $\bm{\alpha}$, $\bm{\mu}$ and $\bm{\Sigma}$. Whereas, the posterior probability specified by \eqref{pesos_LCWMG_x} will be used to obtain values of $\bm{Z}$ for individuals with missing data, specifically used in the imputation step included in the algorithm.\\

The steps of the implemented procedure are presented in Algorithm~\ref{algo:cwm} and are based on the Gibbs sampler algorithm for the Gaussian FMM \citep{kim2014multiple}. The output variable can be partitioned in the form $\bm{y}=(\bm{y_\text{obs}},\bm{y_\text{mis}})$, where $\bm{y_\text{obs}}$ denotes the observed part and $\bm{y_\text{mis}}$ denotes the missing part. For classification purposes, the notation $\bm{z}$ is used for the variable that classifies the observations in the process of estimating the model parameters, while  $\bm{z_\text{mis}}$ is used to classify the observations with missing information for the imputation process. Two imputation steps are incorporated into the algorithm to update $\bm{z_\text{mis}}$ and $\bm{y_\text{mis}}$.\\

\begin{algorithm}[htp]
\DontPrintSemicolon 
\KwIn{$\bm{y}_{\text{obs}}$, $\bm{X}_{\text{obs}}$, $\bm{X}_{\text{mis}}$}
\KwOut{${\bm{y}}_{\text{mis}}$, $\hat{\bm{\alpha}}$, $\hat{\bm{\mu}}$, $\hat{\bm{\Sigma}}$}
\SetAlgoLined
 initialization: $\bm{y_\text{mis}}^{(0)}$, $\bm{\alpha}^{(0)}$, $\bm{\mu}^{(0)}$, $\bm{\Sigma}^{(0)}$\;
 \For{$j=1,...,J$}{
  generate $\bm{z}^{(j)}$ from 
  $p(\bm{z}|\bm{y_\text{obs}},\bm{X}_{\text{obs}},\bm{X}_{\text{mis}},\bm{y_\text{mis}}^{(j-1)},\bm{\alpha}^{(j-1)},\bm{\mu}^{(j-1)},\bm{\Sigma}^{(j-1)})$\;
  generate $\bm{\nu}^{(j)}$ from 
  $p(\bm{\nu}|\bm{z}^{(j)})$\;
  calculate $\bm{\alpha}^{(j)} = f(\bm{\nu}^{(j)})$\;
  \For{$i=1,...,G$}{
  generate $\Sigma_i^{(j)}$ from 
  $p(\Sigma_i|\bm{y_\text{obs}},\bm{X}_{\text{obs}},\bm{X}_{\text{mis}},\bm{y_\text{mis}}^{(j-1)},\bm{z}^{(j)})$\;
  generate $\bm{\mu}_i^{(j)}$ from 
  $p(\bm{\mu}_i|\Sigma_i^{(j)},\bm{y_\text{obs}},\bm{X}_{\text{obs}},\bm{X}_{\text{mis}},\bm{y_\text{mis}}^{(j-1)},\bm{z}^{(j)})$\;
  }
  generate $\bm{z_\text{mis}}^{(j)}$ from 
  $p(\bm{z_\text{mis}}|\bm{X}_{\text{mis}},\bm{\alpha}^{(j)},\bm{\mu}^{(j)},\bm{\Sigma}^{(j)})$\;
  generate $\bm{y_\text{mis}}^{(j)}$ from 
  $p(\bm{y_\text{mis}}|\bm{X}_{\text{mis}},\bm{z_\text{mis}}^{(j)},\bm{\mu}^{(j)},\bm{\Sigma}^{(j)})$\;
  sort $\bm{\alpha}^{(j)}$ in decreasing order\; 
  reorder $\bm{z}^{(j)}$, $\bm{z_\text{mis}}^{(j)}$, $\bm{\mu}^{(j)}$, $\bm{\Sigma}^{(j)}$ based on 
  the order of $\bm{\alpha}^{(j)}$\;
 }
 \KwResult{Complete imputed database and parameters posterior estimation}
 \caption{\textbf{Gaussian Linear CWM}}
 \label{algo:cwm}
\end{algorithm}

\section{Simulation studies}
\label{sec:estudios_simulacion}

To carry out an analysis of the proposed imputation process, a data set was simulated from a mixture of three-dimensional normal distributions with two components. One of the variables was considered as an output variable, while the other two were considered as input variables. The database contains $n = 1000$ observations of the form $(x_1,x_2,y)$. The mixing probabilities are $\alpha_1=0.6$ and $\alpha_2=0.4$, the mean vectors $\bm{\mu}_1=(1{.}0,9{.}0,7{.}0)$ and $\bm{\mu}_2=(1{.}0,3{.}0,3{.}0)$, and the covariance matrices are
\begin{equation*}
	\Sigma_1=\begin{pmatrix} 
	\phantom{-}1{.}00 & \phantom{-}0{.}50 & \phantom{-}0{.}50 \\ 
	\phantom{-}0{.}50 & \phantom{-}1{.}00 & \phantom{-}0{.}50 \\ 
	\phantom{-}0{.}50 & \phantom{-}0{.}50 & \phantom{-}1{.}00  
	\end{pmatrix} \quad \text{and} \quad \Sigma_2=\begin{pmatrix} 
	\phantom{-}1{.}00 & \phantom{-}0{.}50 & -0{.}50 \\ 
	\phantom{-}0{.}50 & \phantom{-}1{.}00 & -0{.}50 \\ 
	-0{.}50 & -0{.}50 & \phantom{-}1{.}00  
	\end{pmatrix}.
\end{equation*}
For cluster 1, $50\%$ of the data was randomly selected and considered missing, while only $10\%$ was selected for cluster 2. A summary of how the data was generated is presented in Table \ref{cuadro:datos_simulados}. The table shows the corresponding count of observations, in addition to the percentages by rows and by columns. In the complete data column, it is observed that 57.5\% were simulated for the first cluster and 42.5\% for the second. The missing column shows that 51.1\% of the data were generated as missing for cluster 1 while 11.3\% for cluster 2. While the information observed by cluster was 42.7\% for cluster 1 and 57.3\% for cluster 2. Additional information can be consulted in the table. 
\begin{table}[h]
\centering
\resizebox{11.5cm}{!}{
\begin{tabular}{c|cl|cl|cl}
\multicolumn{1}{l|}{} & \multicolumn{2}{c|}{\textbf{observed}} & \multicolumn{2}{c|}{\textbf{missing}} & \multicolumn{2}{c}{\textbf{complete}}  \\ 
\hline		
\multirow{2}{*}{\textbf{cluster 1}}  & 281 & \textit{(48.9\%)} & 294   & \textit{(51.1\%)} & 575 & \textit{(100\%)} \\
& \multicolumn{1}{l}{\textit{(42.7\%)}} &  & \multicolumn{1}{l}{\textit{(86.0\%)}} & & \multicolumn{1}{l}{\textit{(57.5\%)}} & \\ 
\hline
\multirow{2}{*}{\textbf{cluster 2}} & 377  & \textit{(88.7\%)} & \phantom{-}48  & \textit{(11.3\%)} & 425 & \textit{(100\%)} \\
& \multicolumn{1}{l}{\textit{(57.3\%)}} &  & \multicolumn{1}{l}{\textit{(14.0\%)}} &        & \multicolumn{1}{l}{\textit{(42.5\%)}} & \\ 
\hline
\multirow{2}{*}{\textbf{total}} & 658 & \textit{(68.3\%)} & 342 & \textit{(31.7\%)} & 1000   & \textit{(100\%)} \\
& \multicolumn{1}{l}{\textit{(100\%)}}  &  & \multicolumn{1}{l}{\textit{(100\%)}}  & & \multicolumn{1}{l}{\textit{(100\%)}}  &                 
\end{tabular}
}
\caption{Distribution of observed, missing and complete data by cluster.}
\label{cuadro:datos_simulados}
\end{table}

\begin{figure}[htb]		
	\centering
    \subcaptionbox{Projection $X_1 \times Y$.	\label{fig:datos_proyecciones_1a}}{\includegraphics[scale=0.235]{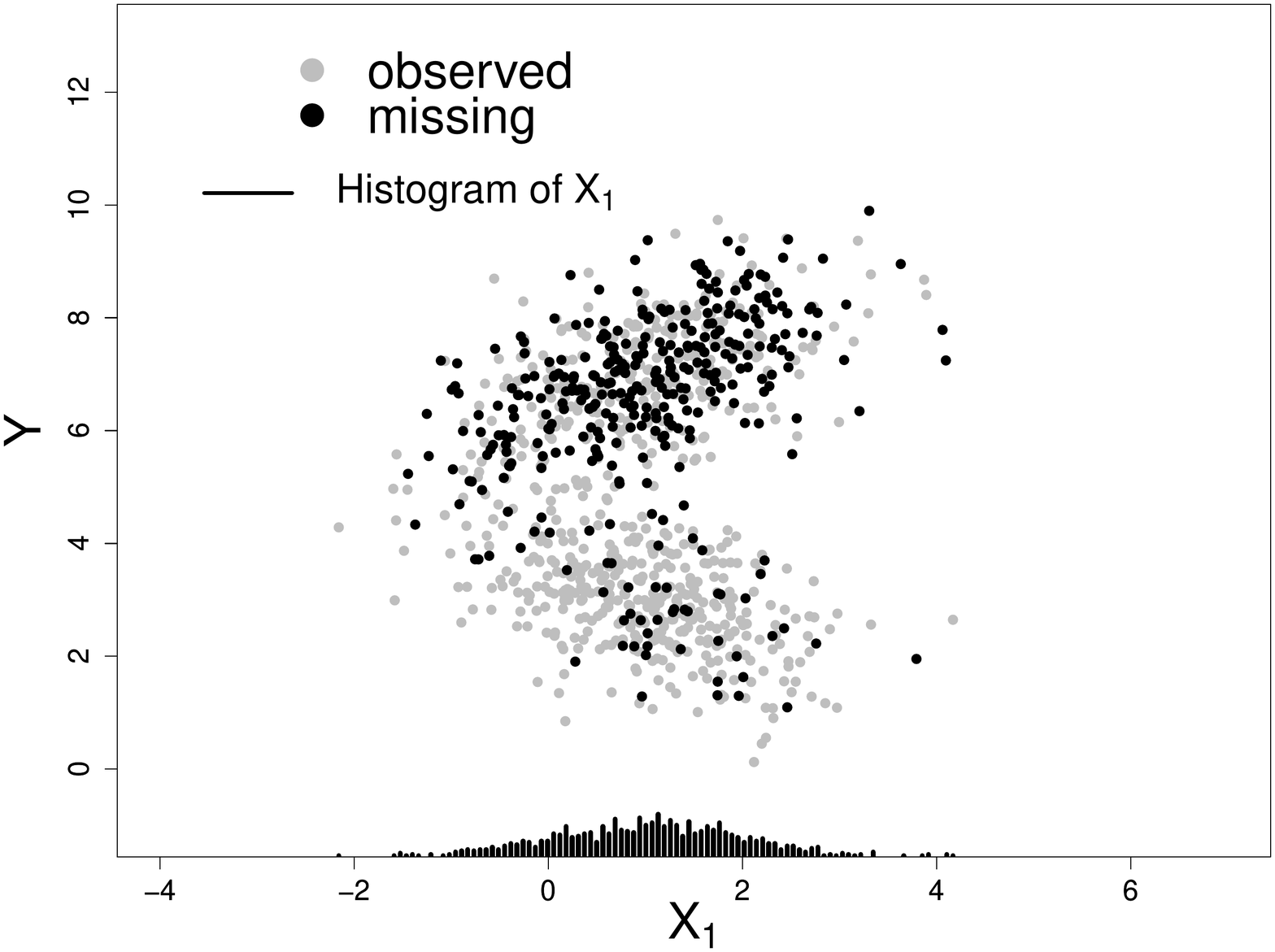}}		
    \subcaptionbox{Projection $X_2\times Y$. \label{fig:datos_proyecciones_1b}}{\includegraphics[scale=0.235]{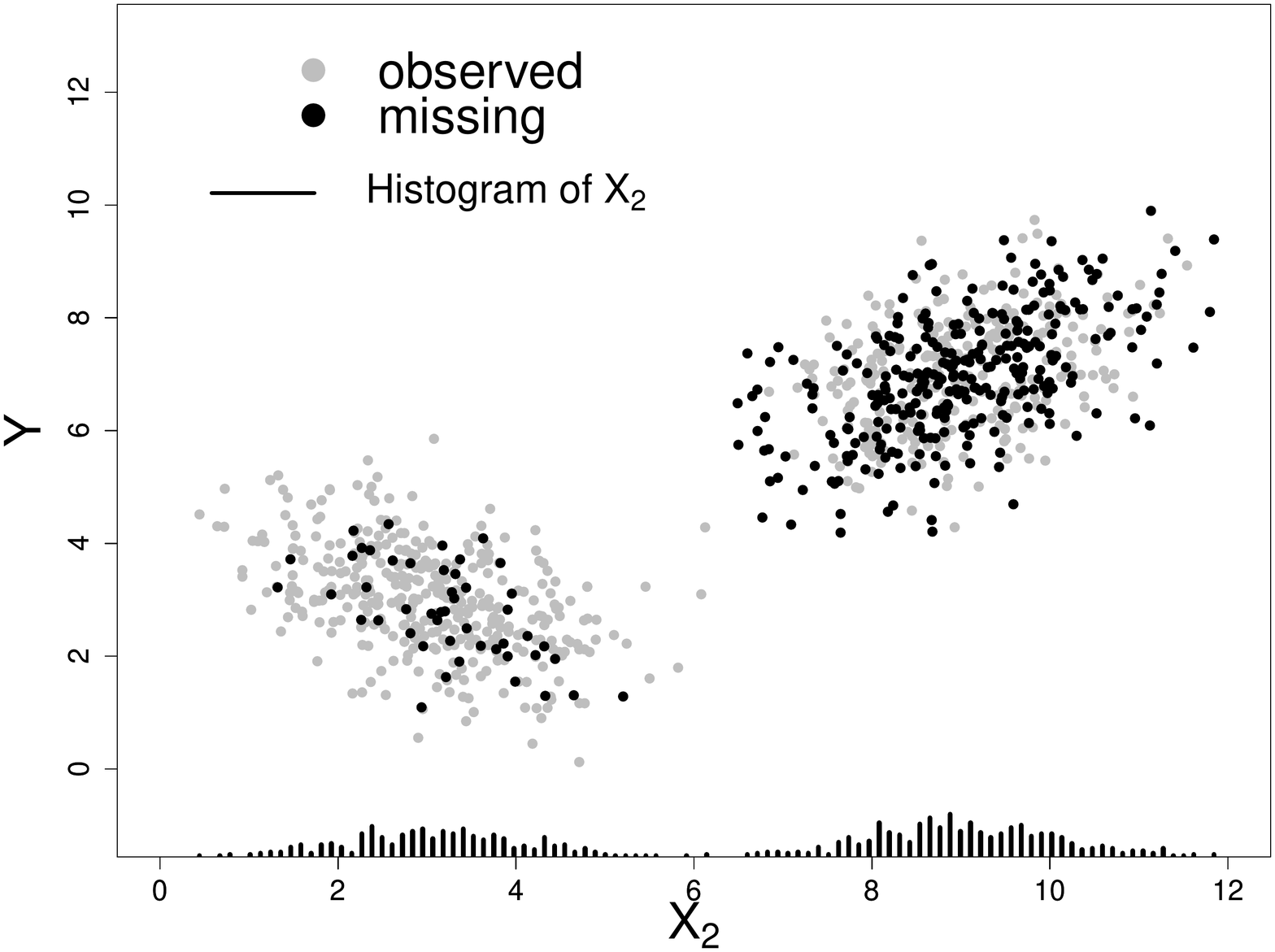}}		
	\caption{Scatter plots for observed and missing data for the simulated database.}	
	\label{fig:datos_proyecciones_1}
\end{figure}

Scatter plots of the observed and missing data are illustrated in Figure \ref{fig:datos_proyecciones_1}, with the projection plane $X_1 \times Y$ in Figure \ref{fig:datos_proyecciones_1a} and the projection plane $X_2 \times Y$ in Figure \ref{fig:datos_proyecciones_1b}. Since the probability of missing is the same within each component, considering the variable $Y$ as the one with missing information and the variables $X_1$ and $X_2$ as fully observed, the missing data mechanism can be assumed as Missing at Random (MAR), \citep{van2018flexible, rubin1976inference}.\\

In Section~\ref{subsec:simulacion1}, we studied how the input variable can influence the imputation process in our model. Two scenarios characterized by input variables with extreme behaviors are analyzed. An additional scenario is presented where the data pattern is generated from an MNAR mechanism on one of the components. In Section~\ref{subsec:simulacion2} we compare our imputation model with other imputation methods proposed in the missing data literature.\\

\subsection{Model performance with new information}
\label{subsec:simulacion1}
Two scenarios are analyzed from the simulated data. In both, the variable $Y$ will be treated as the output variable and the one with missing information. The variables $X_1$ and $X_2$ will be considered as fully observed input variables and provide auxiliary information for the imputation processes. In the first scenario (\textit{Scenario 1}), Figure \ref{fig:datos_proyecciones_1a}, the model imputes the variable $Y$ with the information from the variable $X_1$. The information provided by $X_1$ does not allow to conclude with which of the two components to impute. For the second scenario (\textit{Scenario 2}), the variable $Y$ is imputed with information from the variable $X_2$. Figure \ref{fig:datos_proyecciones_1b} allows us to conclude that, knowing information on this variable, it is possible to decide correctly which component to impute from.\\

The histograms at the bottom of the two graphs in Figure \ref{fig:datos_proyecciones_1}, and which refer to the distributions of the input variables in each case, allow us to observe the difference in their behavior. In the case of the histogram on the right side, where two totally separate groups are observed for the distribution of $X_2$, it will indicate that the input variable is distributed \textit{separately among components}, this is an ideal behavior for the imputation process.\\

In all simulated scenarios, the imputation program was run on \textsf{R} software \citep{R} maintaining similar conditions with {\ttfamily burn-in = 10000} and a sample size adjusted for autocorrelation, {\ttfamily effectiveSize=1500}, implemented using {\ttfamily coda} package \citep{coda}. The number of components was established at the fixed value of {\ttfamily G=10}. In all cases, only the first two components were occupied in the fitted models. The trace plots performed well, guaranteeing the convergence of the chains. To summarize the classification and imputation process, 
the iteration that maximizes the density \textit{a posteriori} (MAP) is chosen \citep{fraley2007bayesian}. Thus, all the graphs, the estimates, and the general descriptions provided here are based on the results obtained at this iteration. For the graphs corresponding to the posterior probabilities that the observation $\bm{x}$ belongs to the component $Z$, the notation $\alpha_Z(\bm{x})=p(Z|\bm{x})$ is used, following the idea of the notation for mixing probabilities, $\alpha_Z=p(Z)$.\\

\subsubsection{Scenario 1: A variable with low performance}

In the first scenario, the imputation model is implemented, seeking to complete the missing information for the output variable $Y$, using only the input variable $X_1$ as auxiliary information. As previously discussed, this input variable does not provide information about which component an observation belongs to, thus, special interest is in the behavior of the estimates of the mixing probabilities. For cluster 1, $\hat{\alpha}_1=0{.}418$, while for cluster 2, $\hat{\alpha}_2=0{.}582$. These estimates are strongly influenced by the proportion of data observed in each cluster, and determine the proportion of data imputed in each group, ($40.9\%$ of the data were imputed in cluster 1, while $59.1\%$ were imputed in cluster 2). Likewise, it can be observed how these proportions of imputed data per component are considerably different from the proportions of missing data ($86.0\%$ for cluster 1 and $14.0\%$ for cluster 2), as seen in Table \ref{cuadro:datos_simulados}.\\
\begin{figure}[htp]		
	\centering
	\subcaptionbox{Imputed and missing data. 	\label{fig:datos_proyecciones_2a}}{\includegraphics[scale=0.170]{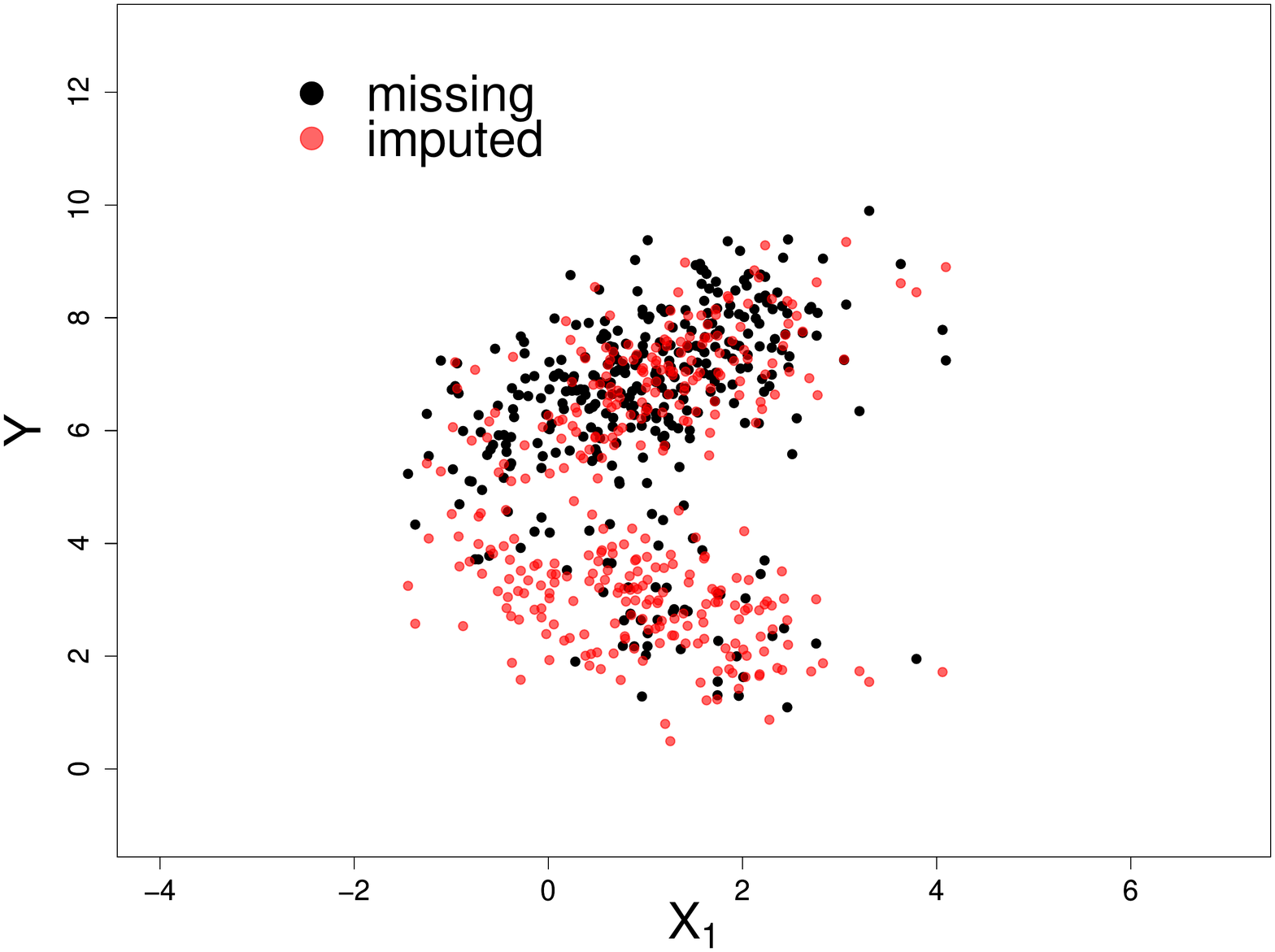}}		
	\subcaptionbox{Imputed and observed data. \label{fig:datos_proyecciones_2b}}{\includegraphics[scale=0.170]{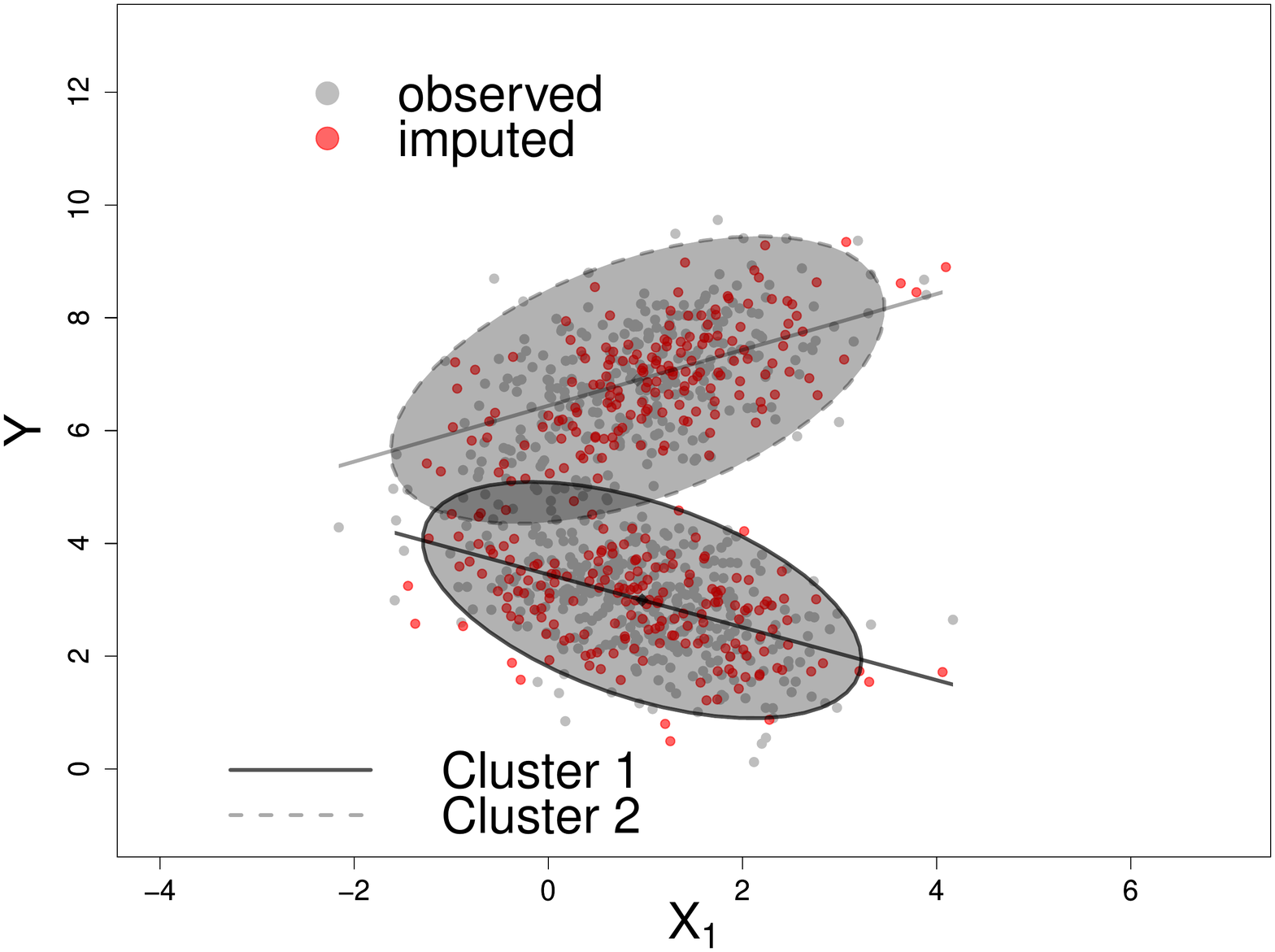}}
	\subcaptionbox{Posterior probability dependent on $\bm{x}_1$ for each cluster. \label{fig:datos_proyecciones_2c}}{\includegraphics[scale=0.170]{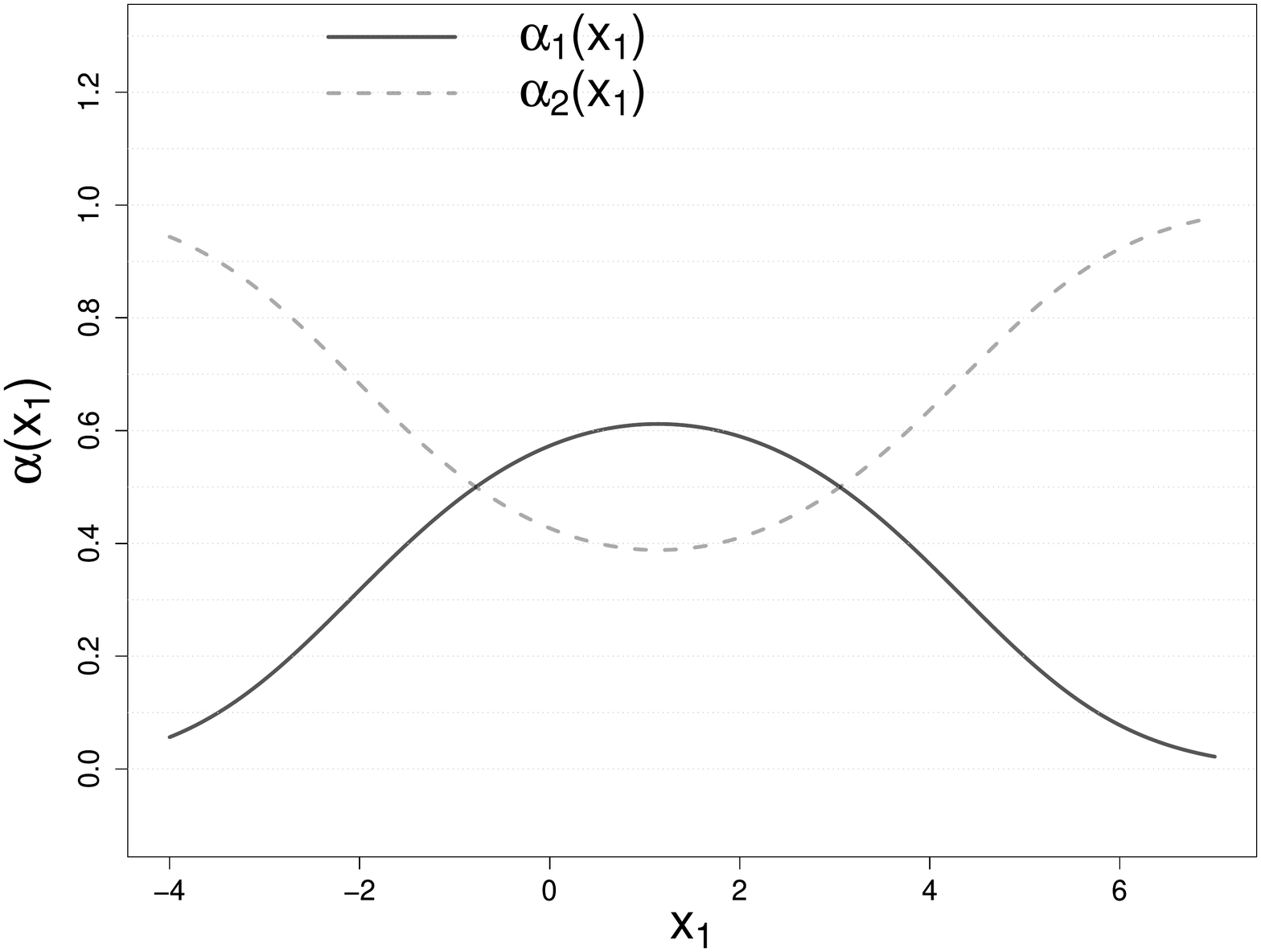}}		
	\caption{Construction of the imputation model in the case of auxiliary information given by the variable $X_1$.}	
	\label{fig:datos_proyecciones_2}
\end{figure}

Figure \ref{fig:datos_proyecciones_2} shows how the imputation model was built and how the missing data were imputed. Figure \ref{fig:datos_proyecciones_2a} presents a scatter plot where the imputed data is plotted, as well as the data considered missing. It can be observed that, although the data is imputed around the centers of each component, the proportion in each differs from the proportion the missing data was generated. Figure \ref{fig:datos_proyecciones_2b} illustrates how the imputation model was built, and presents the observed data that are the basis for the construction of the regression lines. Together with these points, quantile ellipses of $95\%$ allow each of the components to be distinguished; this same graph shows the imputed values around the lines. In particular, the regression lines are the result of the so-called conditional distributions, and are used as predictive models in the imputation process. The marginal distributions are used for the classification process and are the basis for the construction of the curves in Figure~\ref{fig:datos_proyecciones_2c}. In this, the graphs of the posterior probabilities dependent on the input variable $\bm{x}_1$ as defined by the expression \eqref{pesos_LCWMG_x} are shown. For values of $\bm{x}_1$ close to the estimates of the means in the two components ($\hat{\mu}_1= 0.897$, $\hat{\mu}_2= 1.109$), the posterior probabilities are strongly influenced by the proportion of observed data.\\

\subsubsection{Scenario 2: A variable with high performance} \label{sec:scenario2}

In the second case, we use the input variable $X_2$ to perform the imputation process. In this case,
the model returns proportions of imputed data similar to how the missing data were generated. The estimates for the mixing probabilities are $\hat{\alpha}_1=0{.}583$ and $\hat{\alpha}_2=0{.}417$ respectively. The fact that the input variable is capable of separating the components allows the information provided by it to precisely determine which component to impute from. The proportion of imputed data in each group was $14.0\%$ for component 1 and $86.0\%$ for component 2, the same values presented in Table~\ref{cuadro:datos_simulados}.\\ 

\begin{figure}[htb]		
	\centering
	\subcaptionbox{Imputed and missing data. 	\label{fig:datos_proyecciones_3a}}{\includegraphics[scale=0.170]{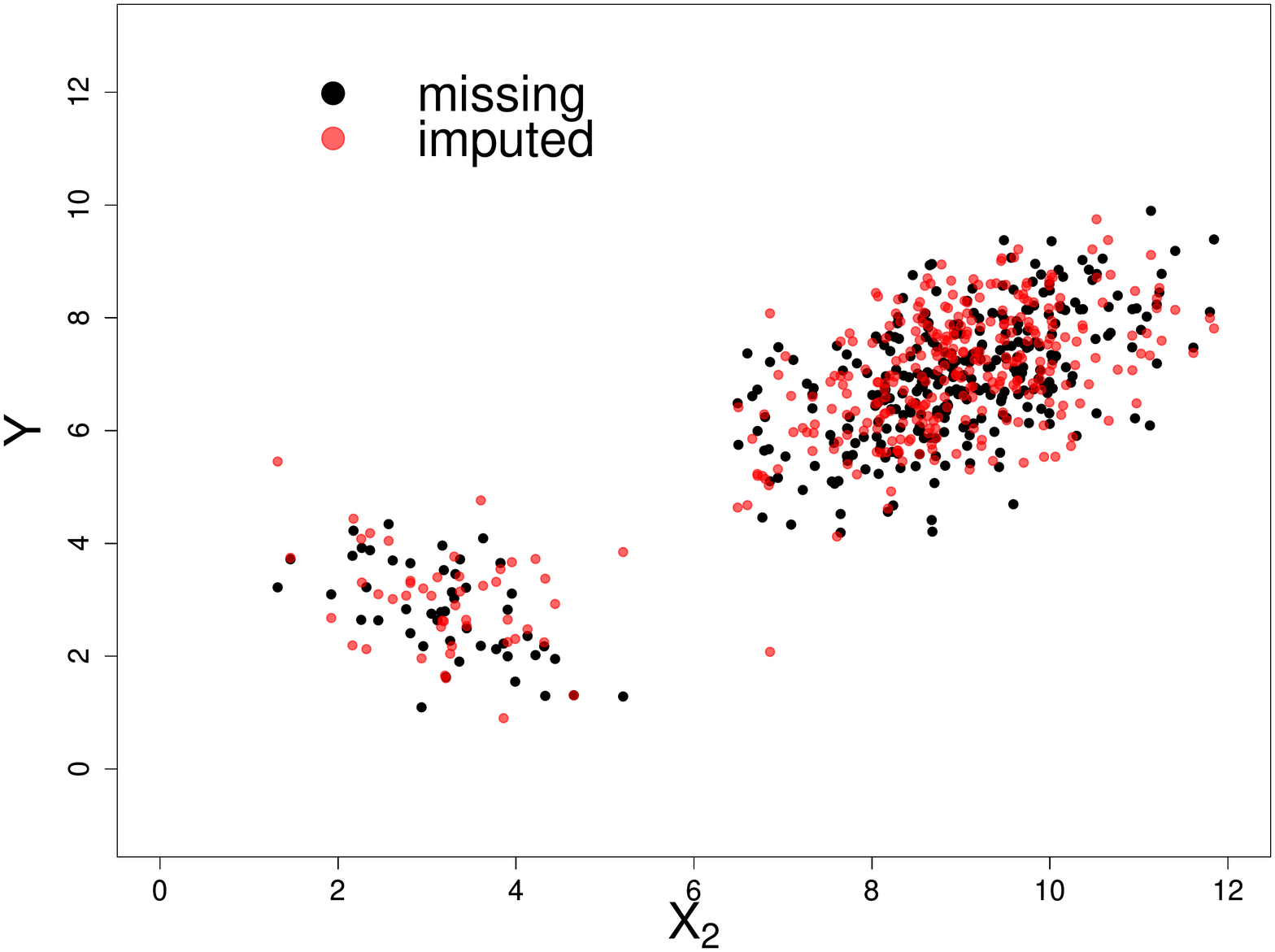}}	
	\subcaptionbox{Imputed and observed data. \label{fig:datos_proyecciones_3b}}{\includegraphics[scale=0.170]{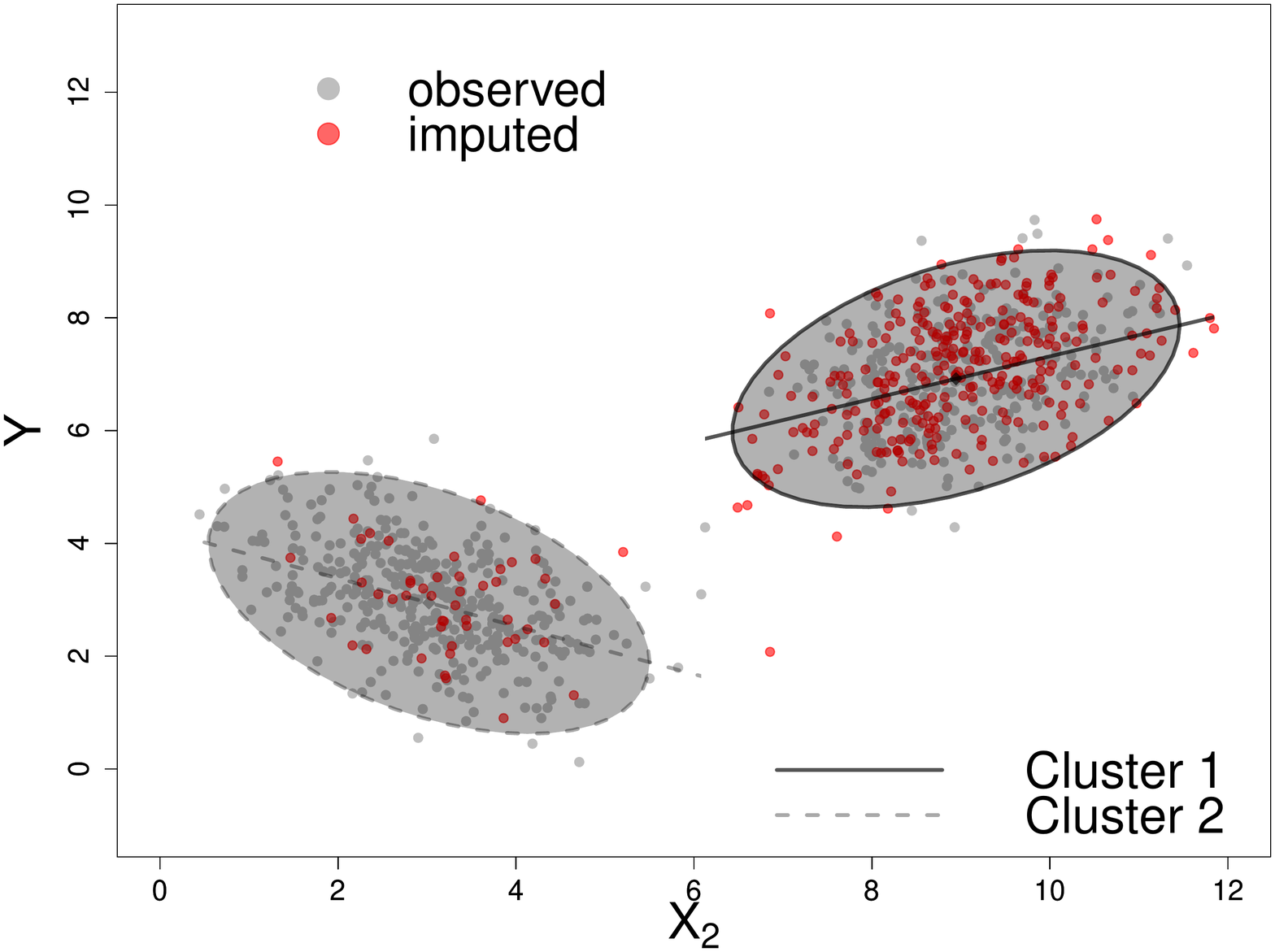}}
	\subcaptionbox{Posterior probability dependent on $\bm{x}_2$ for each cluster. \label{fig:datos_proyecciones_3c}}{\includegraphics[scale=0.170]{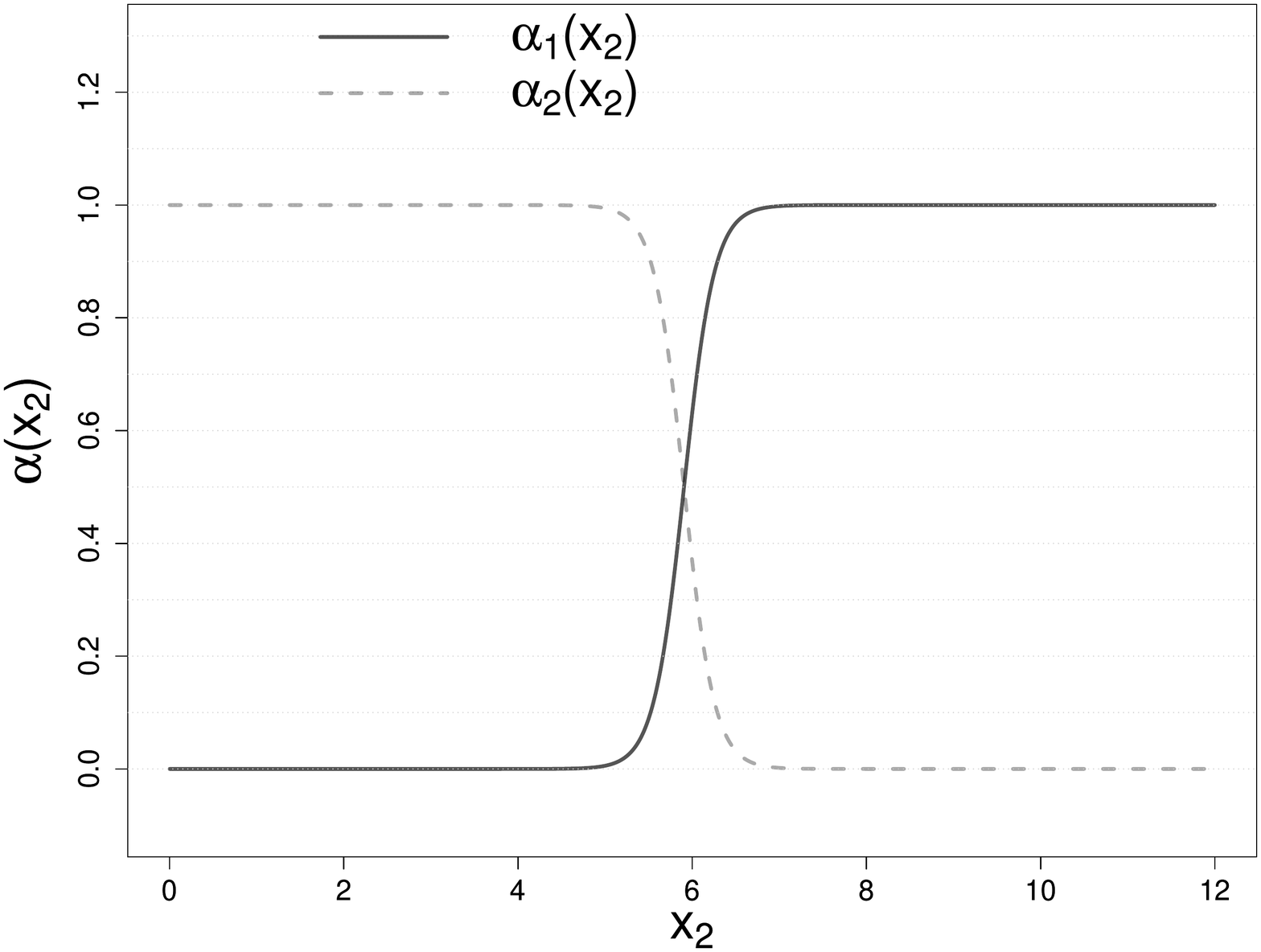}}		
	\caption{Construction of the imputation model in the case of auxiliary information given by the variable $X_2$.}	
	\label{fig:datos_proyecciones_3}
\end{figure}
Figure~\ref{fig:datos_proyecciones_3} shows, for scenario 2, how the data were imputed. 
Figure \ref{fig:datos_proyecciones_3a} shows the data that were considered missing and the imputations made by the model. It is evidenced that, in addition to the imputations being closer to the missing data, the proportions in the two components corresponding to missing data and imputed data are the same. Similar to scenario 1, Figure~\ref{fig:datos_proyecciones_3b} presents how the model was built based on the observed data set, and also illustrates the distribution of the imputed data. Specifically, the conditional and marginal distributions are responsible for carrying out the imputation and classification processes. The regression lines, result of the conditional distributions, are plotted for each component and show, next to the imputed data, the pattern followed to carry out the imputation. Also presented in this graph are $95\%$ confidence ellipses for each component. To complement, and as a product of the marginal distributions together with the mixing probabilities, the posterior probability curves are illustrated in the graph of Figure~\ref{fig:datos_proyecciones_3c}. This graph reflects an ideal behavior regarding the classification process. Punctually, it makes the correct decision regarding which component to impute from, given the value of the input variable. For values of the input variable $X_2$ less than six, the model imputes with probability one from component 1. Likewise, from some value greater than six, the observation is classified in component 2 and imputed with the estimated model for this case. For the values of the input variable around six, there is a transition in the probabilities, in such a way that in the limit, the probabilities to classify in one or another component coincides with the value $0.5$.\\

The scenarios presented illustrate two extreme cases. In the first case, in which the input variable does not provide information about the component to be imputed, the classification proportion remains similar to the observed proportion of the data in each component.
The second case is an ideal scenario for the 
imputation model, where the information provided by the input variable allows us to correctly determine 
which component to impute from.\\

An additional scenario 
considers the imputation process under the joint information of the two input variables treated in the previous scenarios. In this situation, the two variables are integrated into an input vector of the form $\bm{X}=(X_1,X_2)$. As expected, this scenario performs equivalently to the case of scenario 2. In summary, the model identifies that $X_2$ is capable of separating the components while the noise added by $X_1$ is not relevant in the imputation step. The detailed results of this scenario can be consulted in Appendix \ref{apendice_A} and are equivalent to the ones obtained in Scenario 2.\\

\subsubsection{Comparative analysis of simulated scenarios} 

\begin{figure}[htb]
	\includegraphics[width=0.80\textwidth]{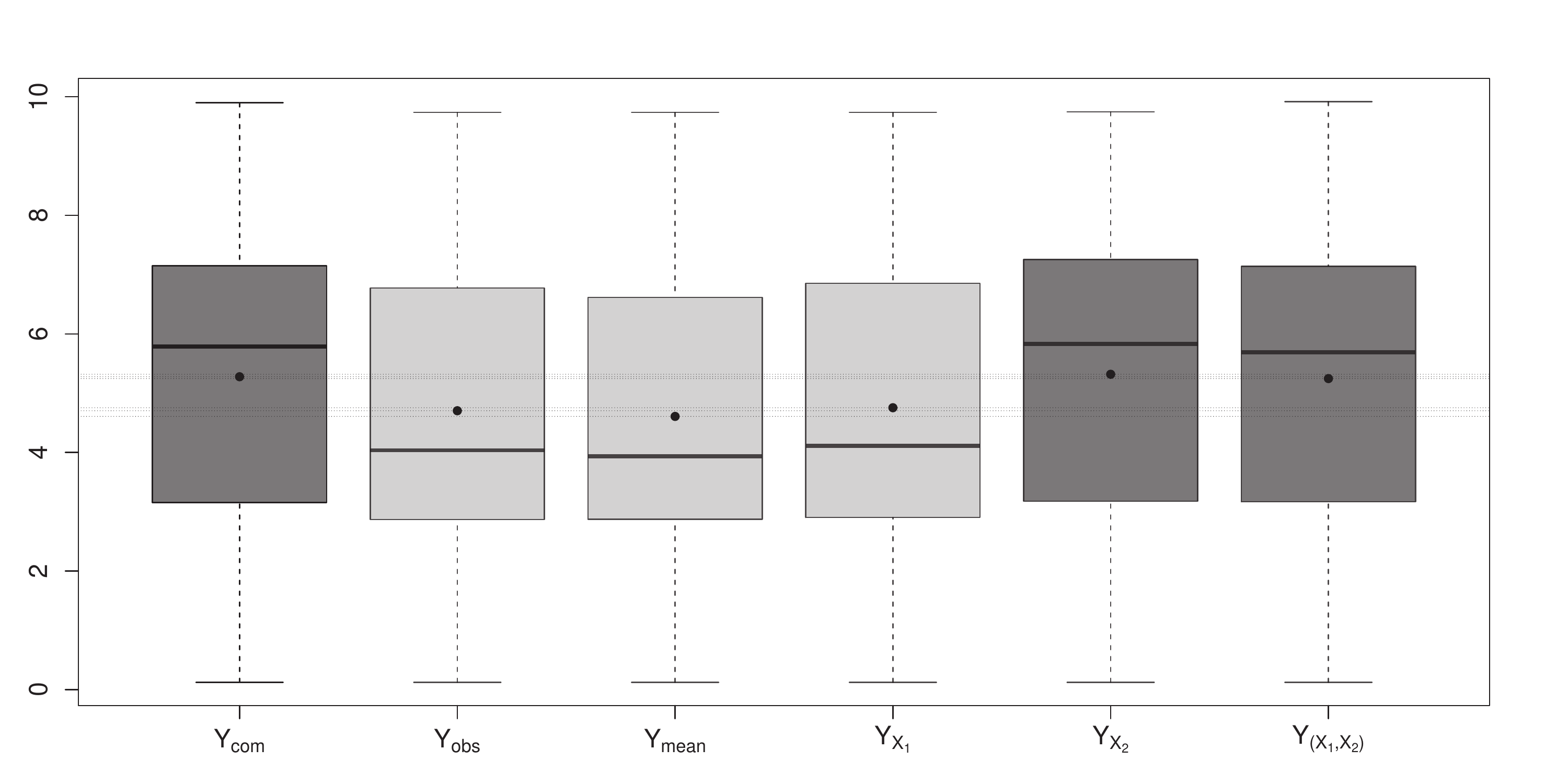}
	\centering
	\caption{Box plots for complete, observed, and the imputed variables with information from the input variables $X_1$, $X_2$, and $(X_1, X_2)$.}
	\label{boxplot}
\end{figure}

Figure~\ref{boxplot} presents box plots for the output variables in the cases of complete information ($\bm{Y}_{\text{com}}$), when only the observed information is considered ($\bm{Y}_{\text{obs}}$) and after the imputation processes ($\bm{Y}_{(\cdot)}$). 
Within each boxplot, the point inside represents the value of the sample mean for each data set.

In the case of the imputation process using the variable $X_1$ (Scenario 1), since the variable does not provide any information about which component to impute from, the model selects said component based on the estimated mixing probabilities. That is the reason for the similarity between the distribution of $\bm{Y}_{\text{obs}}$ and the imputed output variable $\bm{Y}_{X_1}$. 
When we use the $ X_2 $ variable as an input variable in the model (Scenario 2), since the information it provides allows us to identify the component to which an observation belongs, with the imputed output variable $\bm{Y}_{X_2}$ we have a closer proximity to the distribution of the original complete variable $\bm{Y}_{\text{com}}$. Something similar happens with the output variable imputed from the information of the vector $(X_1,X_2)$, denoted as $\bm{Y}_{(X_1,X_2)}$.\\

After a visual interpretation of the imputation capacity of the method, it is desirable to include a quantitative diagnosis to measure the quality of the imputation process. To do so, 
we used the Kullback-Liebler divergence \citep{kullback1951information}. 
For two density functions $f(\cdot)$ and $g(\cdot)$, 
the Kullback-Liebler divergence is defined by the integral
\begin{equation}
\text{KL}(f,g):=\int_{-\infty}^{{\infty}}f(x)\log\frac{f(x)}{g(x)}dx.
\label{ecuac:div_KL}
\end{equation}
The $\text{KL}(f,g)$ divergence can be interpreted as the amount of information lost when we want to approximate the $f$ distribution using the $g$ distribution. Unfortunately, for the case where $f(\cdot)$ and $g(\cdot)$ are Gaussian mixture models, the expression in \eqref{ecuac:div_KL} is analytically intractable. \citet{hershey2007approximating,durrieu2012lower} present several approximations, as well as bounds for the divergence in this case.\\

Thus, to implement the calculation of the KL divergence, the {\ttfamily integrate} function is used in the \textsf{R} software, which allows to calculate integrals of one-dimensional functions over infinite intervals. In this case, it will be used to evaluate the integral in \eqref{ecuac:div_KL} and will be denoted as $\text{KL}_{\text{int}}$. To measure the quality of the imputation process, we calculate a quantile interval of $95\%$ for the KL divergence of {\ttfamily N=10000} replications of the complete dataset obtained randomly from the original distribution using the {\ttfamily mixsmsn} package \citep{prates2013mixsmsn}.\\ 

Any value of KL of a sample data set that is within this interval will allow to conclude that such sample has the characteristics of a original sample from the original distribution. The interval obtained with the complete data replications and the KL divergences for the different imputation processes are presented in Table~\ref{Cuadro:KL_bases_imputadas_op1}. The table also presents the relative distances to the extreme right of the interval, which gives an idea of how far the sample of interest is from the ones of the original distribution. If the value of the KL divergence falls within the interval, it is noted with {\ttfamily WI} (\textit{withing the interval}). In expression \eqref{ecuac:div_KL}, the function $f$ will refer to the original distribution, that is, the one with which the database was generated, while the function $g$ will refer to the distribution estimated from the data set $Y_{(\cdot)}$, specifically noted by $g_{Y_{(\cdot)}}$.\\

\begin{table}[htb]
\centering	
\begin{tabular}{llcclcc}
\cline{3-4}
\multicolumn{1}{l}{} &  & \multicolumn{2}{c}{\textbf{Approach method}}\\ \cline{3-4} 
\multicolumn{1}{l}{} &  & \multicolumn{1}{l}{$\text{KL}_{\text{int}}$} & \multicolumn{1}{l}{Relative distance}  \\ 
\cline{1-1} \cline{3-4}
Qu.int. 95.0\% &  & (0, 0.0055) & - \\ 
\cline{1-1} \cline{3-4} 
\phantom{-----}$\bm{g}_{\bm{Y}_{\text{com}}}$ &  & 0.0029 & \phantom{-}{\ttfamily WI}\\
\phantom{-----}$\bm{g}_{\bm{Y}_{\text{obs}}}$  &  & 0.0679 & 12.25\\
\phantom{-----}$\bm{g}_{\bm{Y}_{X_1}}$         &  & 0.0665 & 12.00\\
\phantom{-----}$\bm{g}_{\bm{Y}_{X_2}}$         &  & 0.0034 &  \phantom{-}{\ttfamily WI} \\
\phantom{-----}$\bm{g}_{\bm{Y}_{(X_1, X_2)}}$  &  & 0.0036 & \phantom{-}{\ttfamily WI}\\ 
\cline{1-1} \cline{3-4} 
\end{tabular}	
\caption{KL divergences and relative distances for the imputed variables with information from the input variables $X_1$, $X_2$, and $(X_1, X_2)$.}
\label{Cuadro:KL_bases_imputadas_op1}	
\end{table}

For $\bm{Y}_{\text{com}}$, as expected, it is observed that the KL value is within the quantile interval of $95\%$. For the distribution of the observed data, $\bm{Y}_{\text{obs}}$, a relative distance of approximately twelve units from the samples of the original distribution is observed. The samples of the imputed variable 
$\bm{Y}_{X_1}$ have similar results to those from the observed data set. 
The imputation using variables $\bm{Y}_{X_2}$ and $\bm{Y}_{(X_1, X_2)}$ have the best behaviors, both recover samples from the original distribution. In the case of the imputation process with the vector $(X_1,X_2)$, the inclusion of a meaningful variable that is distributed separately among components, as is the case of the input variable $X_2$, allows for the whole vector to be distributed separately among components, justifying the results obtained.\\

\begin{figure}[htb]
\centering
\includegraphics[width=0.65\textwidth]{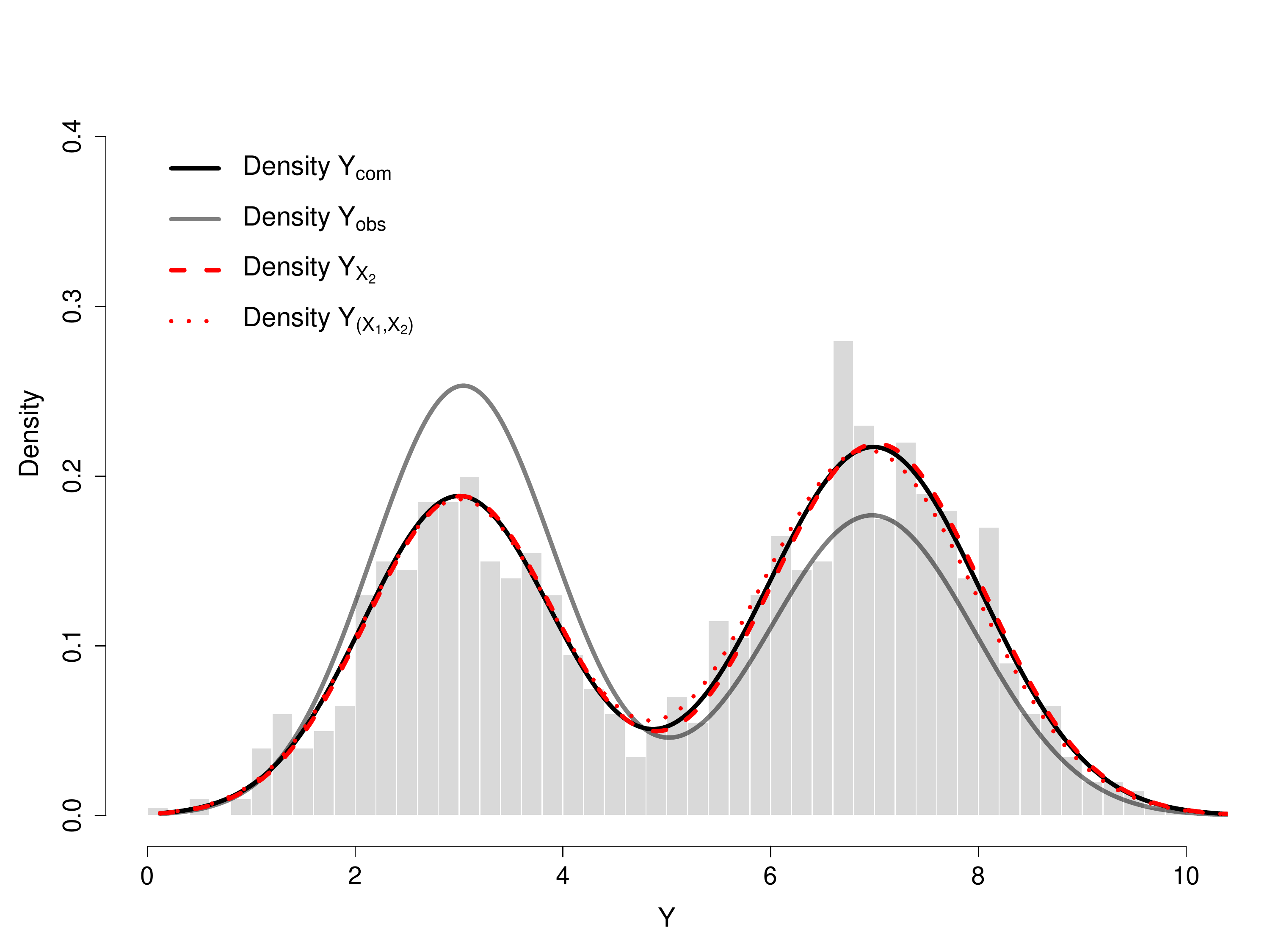} 
\caption[Estimated densities for the imputed variables with information from the input variables.]{Histogram of the variable $\bm{Y}_\text{com}$ and estimated densities for the imputed variables with information from the input variables $X_1$, $X_2$, and $(X_1, X_2)$.}
\label{fig:histdens_KLopc1}
\end{figure}

The same estimates obtained from the imputed variables and used to calculate the KL divergences are used to obtain the graphs of the estimated densities shown in Figure \ref{fig:histdens_KLopc1}. A histogram is presented illustrating the distribution of the complete variable, $\bm{Y}_\text{com}$. The solid black and gray curves represent the estimated densities for $\bm{Y}_\text{com}$ and $\bm{Y}_\text{obs}$, respectively. Once again, the good performance of the imputation process can be confirmed with information from the input variable $X_2$ and the input vector $(X_1, X_2)$. The blue dotted curve and the green dashdot curve corresponding to these two estimated densities, respectively, are closest to the estimated density of the variable $\bm{Y}_\text{com}$. This is not the case with the red dashed curve and that corresponds to the variable $\bm{Y}_{X_1}$.\\

In conclusion, the presented imputation methodology makes use of the information provided by the input variables. This information can move between two situations: the input variable does not provide any information about which component the observation belongs to; or the input variable is capable of correctly determining the component to which said observation belongs. Information from several input variables can be taken into account together in the classification process. Therefore, to provide an adequate imputation it is necessary to have inputs that accurately separate the imputation region. If the user does not know which input that is, we suggest the inclusion of an input vector. As shown by our simulations, the performance of the imputation method is not affected by the non-informative inputs thus continuing to offer an appropriate imputation for the data.\\

\subsubsection{A scenario with missing data from a MNAR mechanism}

Missing Not at Random (MNAR) is a missing data mechanism in which there is a relationship between the propensity of a value to be missing and its values. In other words, data are MNAR when the missing values of a variable are related to the values of that variable itself, even after controlling for other variables \citep{rubin1976inference}.\\

\begin{figure}[htb]
\centering
\subcaptionbox{Distribution of observed and missing data.
\label{fig:X1_MNAR_a}}{\includegraphics[scale=0.170]{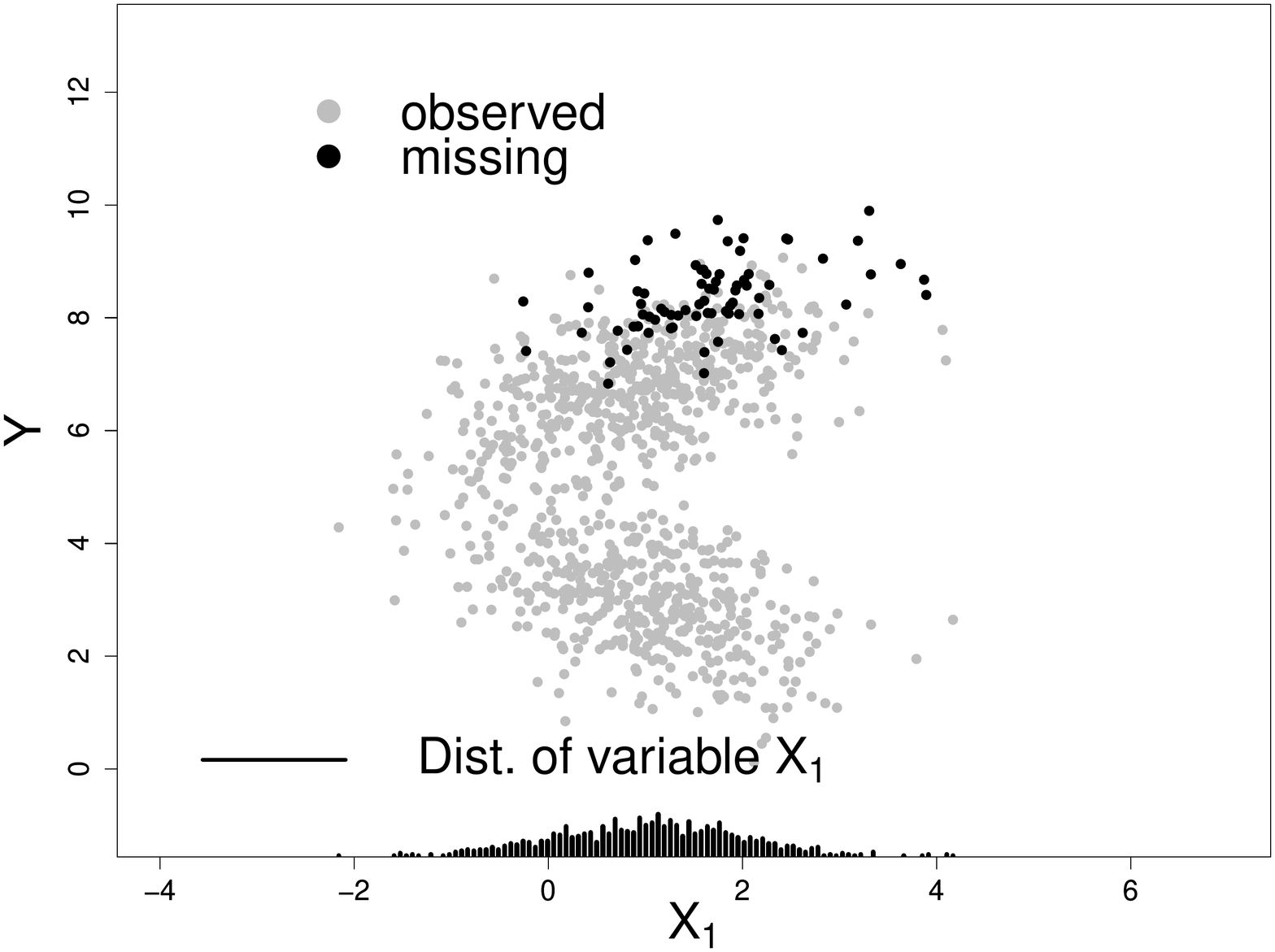}}
\subcaptionbox{Distribution of imputed and missing data.
\label{fig:X1_MNAR_b}}{\includegraphics[scale=0.170]{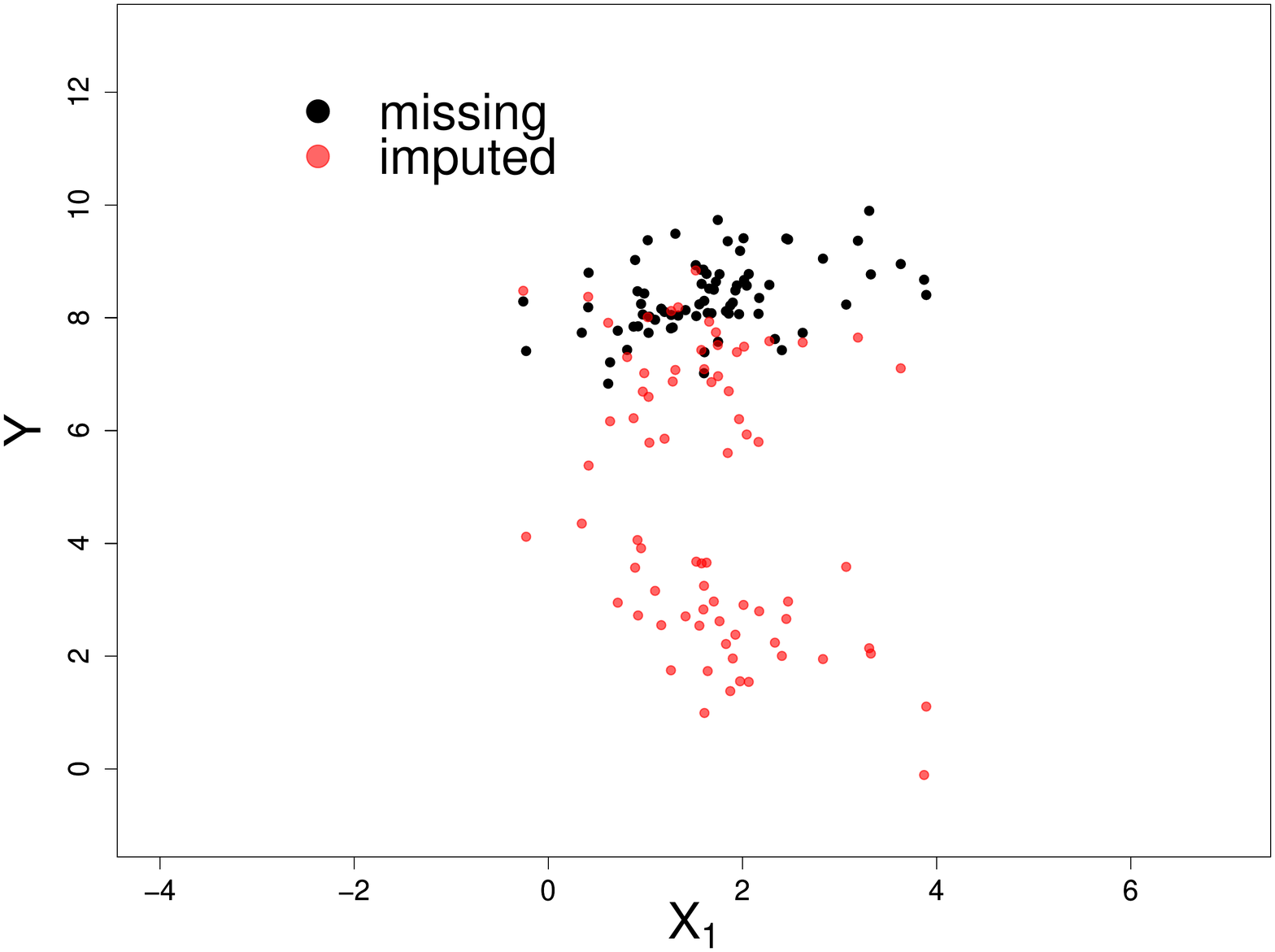}}
\subcaptionbox{Imputation model with information on variable $X_1$.
\label{fig:X1_MNAR_c}}{\includegraphics[scale=0.170]{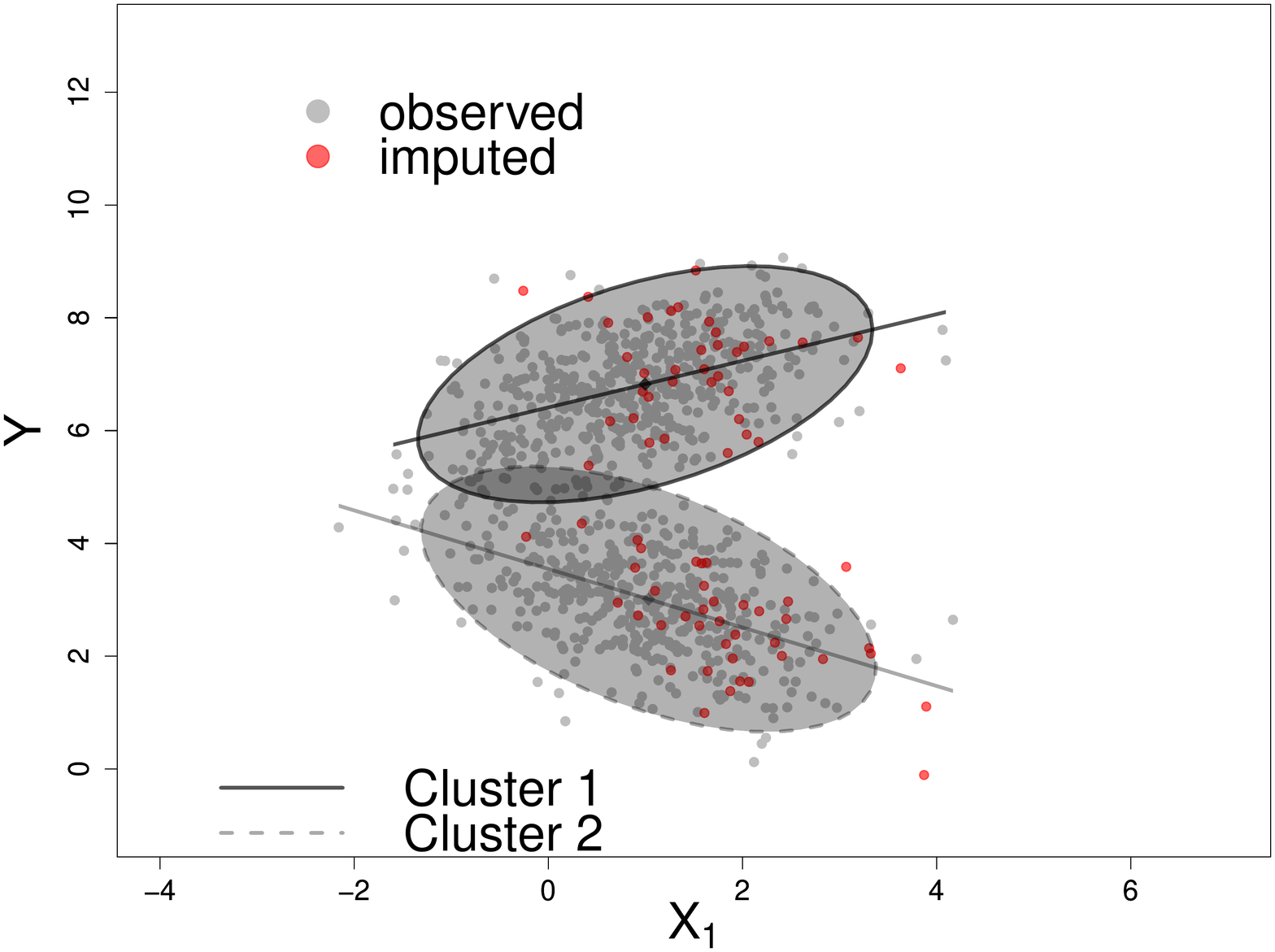}}
\caption[Simulation of a MNAR mechanism and imputation using the variable $X_1$.]{Simulation of a MNAR mechanism and imputation using the variable $X_1$ in the univariate case.}
\label{fig:X1_MNAR}
\end{figure}

\begin{figure}[htb]
\centering
\subcaptionbox{Distribution of observed and missing data.
\label{fig:X2_MNAR_a}}{\includegraphics[scale=0.170]{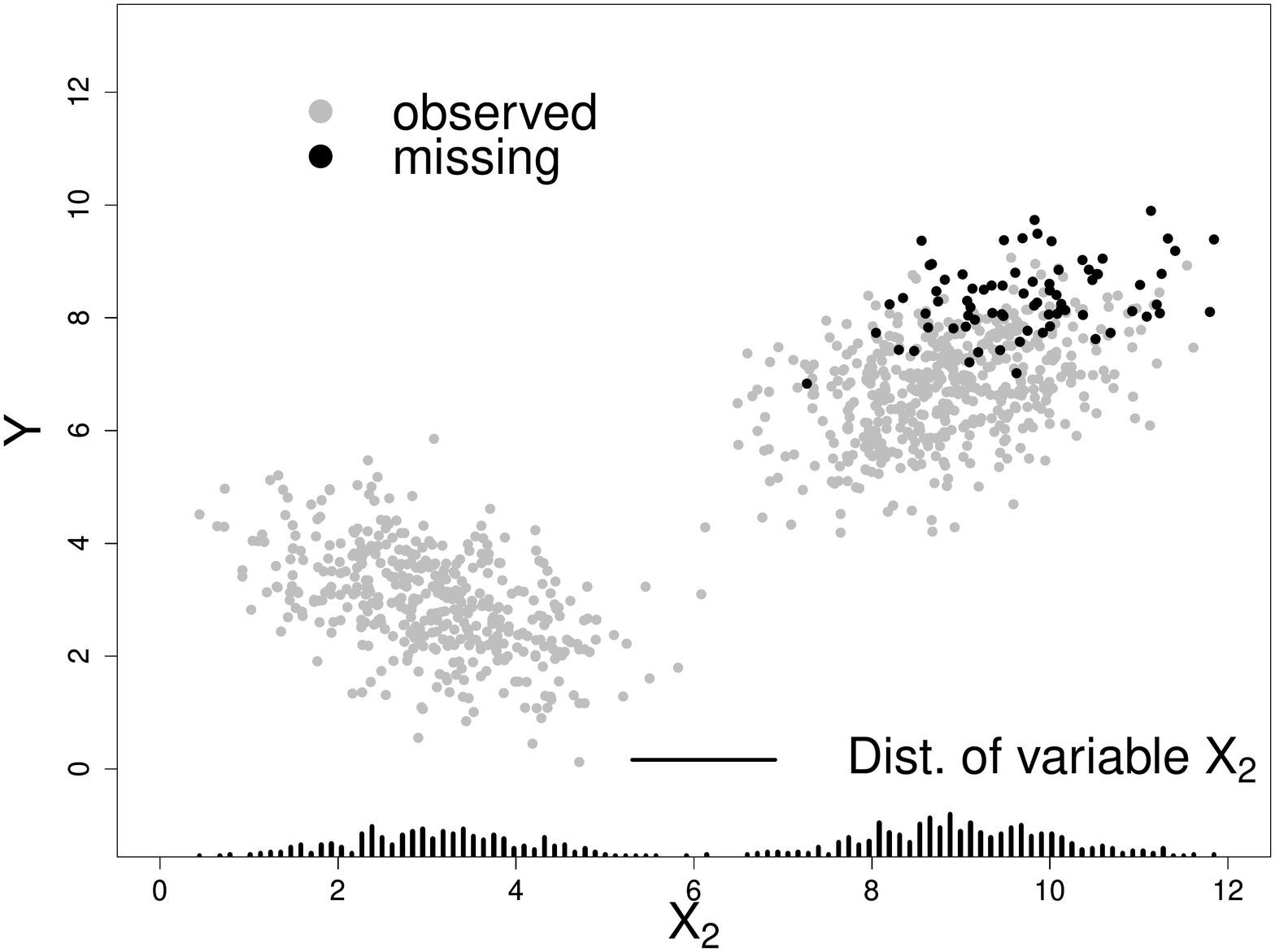}}
\subcaptionbox{Distribution of imputed and missing data.
\label{fig:X2_MNAR_b}}{\includegraphics[scale=0.170]{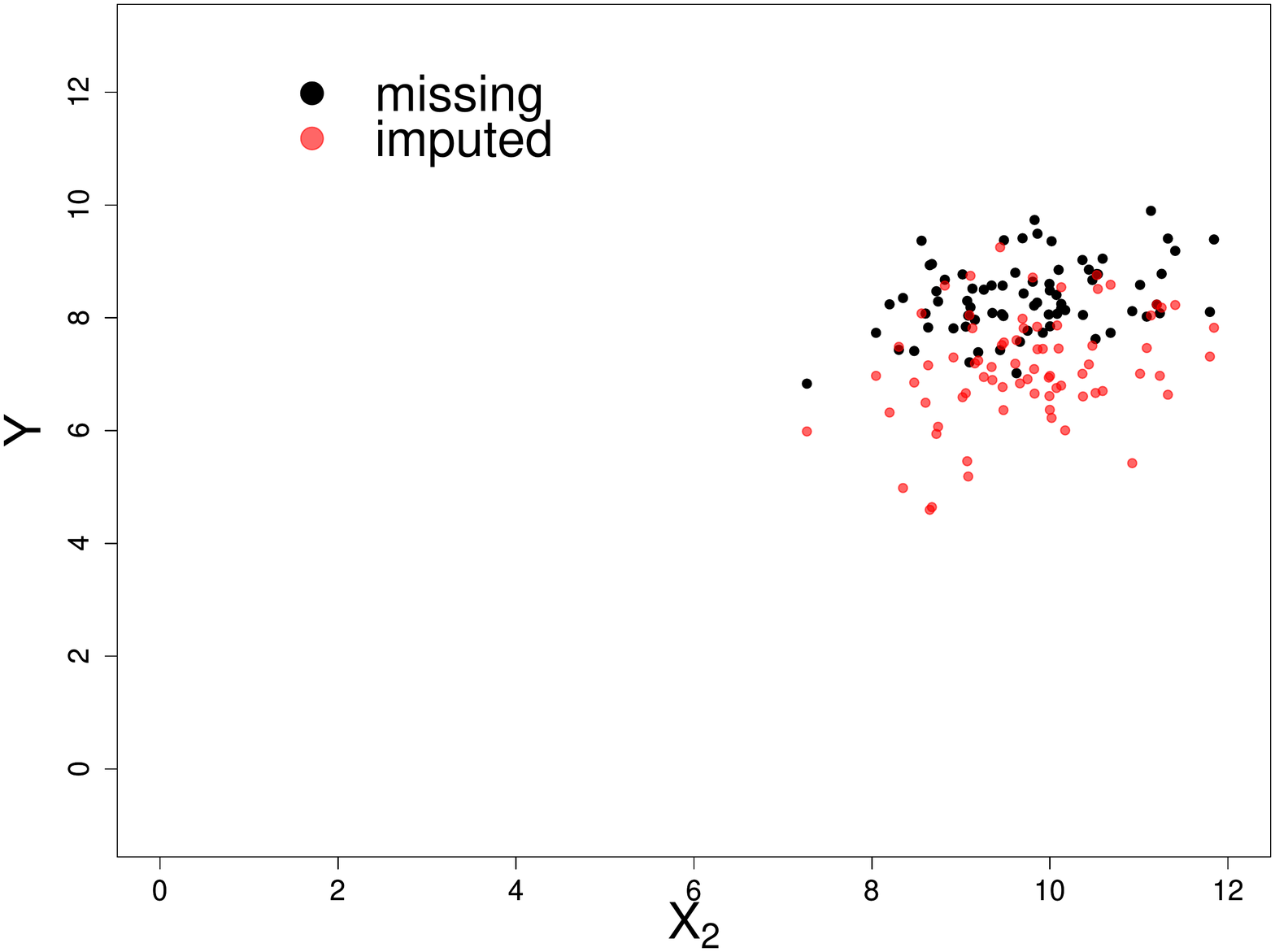}}
\subcaptionbox{Imputation model with information on variable $X_1$.
\label{fig:X2_MNAR_c}}{\includegraphics[scale=0.170]{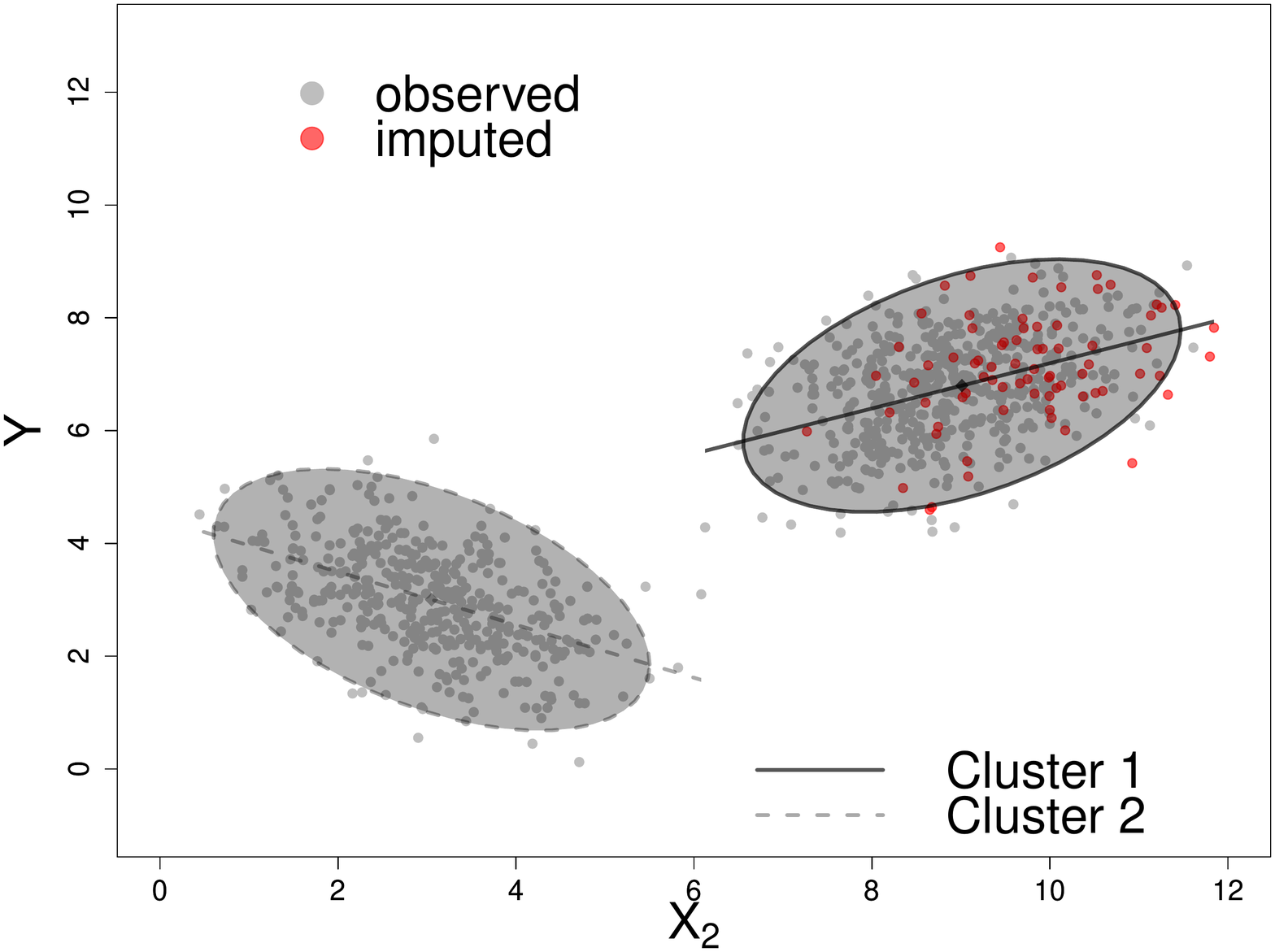}}
\caption[Simulation of a MNAR mechanism and imputation using the variable $X_2$.]{Simulation of a MNAR mechanism and imputation using the variable $X_2$ in the univariate case.}
\label{fig:X2_MNAR}
\end{figure}

On the same dataset, a MNAR mechanism of missing data was simulated to observe how our imputation model behaves. Thus $20\%$ of data with values of $Y$ greater than $6.5$ were randomly selected and considered missing. This procedure generated $75$ missing data points, all belonging to cluster 1. The distribution of the observed and missing data projected on the planes $X_1 \times Y$ and $X_2 \times Y$ can be seen in Figures~\ref{fig:X1_MNAR_a} and \ref{fig:X2_MNAR_a}, respectively. The missing data is grouped at the top of the point cloud corresponding to cluster 1.\\

Similar to scenario 1 presented before, the imputation model was implemented using the input variable $X_1$ as auxiliary information. Since this variable does not offer any information on which component to impute from, the classification procedure is carried out based on the estimates made with the mixing probabilities, $\hat{\alpha}_1=0{.}526$ and $\hat{\alpha}_2=0{.}474$. Thus, $39$ observations are imputed for component 1, while $36$ observations are imputed for component 2, see Figure \ref{fig:X1_MNAR_b}. Furthermore, Figure~\ref{fig:X1_MNAR_c} shows how the imputations are made based on the values of the variable $X_1$ around the regression lines, without considering the region where the missing data is located. Implementing the model with information from the input variable $X_2$ leads to estimates of the mixing probabilities equal to $\hat{\alpha}_1=0.565$ and $\hat{\alpha}_2=0.435$. However, since the input variable $X_2$ provides precise information on which component to impute from, all observations are imputed in component 1, see Figure~\ref{fig:X2_MNAR_b}. Of utmost importance, it should be noted that, although the model imputes accurately in the correct component, it does not have information about the $Y$ region where the missing data came from (Figure~\ref{fig:X2_MNAR_c}). Hence, it can be concluded that, because the model uses a MAR imputation mechanism, it can go up to identifying the correct component but not the exact region within it. To do so, other(s) input(s) variable(s) capable of slicing the $Y$ space accordingly to the missing process are required.\\

\subsection{Gaussian LCWM performance relative to other imputation methods}
\label{subsec:simulacion2}


To compare the proposed imputation method with others, the \textsf{R} packages repository regarding missing data was consulted\footnote{\url{https://CRAN.R-project.org/view=MissingData}}. From our search, three methods were of interest in this process since they are based in the same MAR premises as ours. The first one is the \textit{Bayesian multiple imputation} \citep[{\ttfamily norm},][]{rubin1987multiple} which is a special case of ours when $G=1$; the second method is also a special case and imputes without auxiliary information, in other words, when the components are identified only by their mean \citep[{\ttfamily mean},][]{paiva2014multiple,paiva2017stop}; and the third method is the \textit{predictive mean matching} \citep[{\ttfamily pmm},][]{little1988missing} implemented in the {\ttfamily MICE} package \citep{mice2011} by the function {\ttfamily mice.impute.pmm()}. These methods are imputation procedures that rely on regression models from a Bayesian perspective \citep{van2018flexible}. For our method, the {\ttfamily cwm} notation will be used.\\


The KL divergence values presented in Table~\ref{tabla:cwm_norm_pmm_simulacion1} were obtained by implementing the imputation methods {\ttfamily norm}, {\ttfamily mean} and {\ttfamily pmm} on the simulated data set from Section~\ref{subsec:simulacion1}. From the results obtained, a better performance of the methods {\ttfamily cwm} and {\ttfamily pmm} compared to {\ttfamily norm} and {\ttfamily mean} can be observed. Since the {\ttfamily mean} method does not include inputs, for visualization purposes, it was included in the third line of Table~\ref{tabla:cwm_norm_pmm_simulacion1}. It is clear that this method mimics the observed data set and imputes data following the observed characteristics.\\  
\begin{table}[htb]
\resizebox{15.0cm}{!}{%
\begin{tabular}{llcclcclcc}
\cline{3-4} \cline{6-7} \cline{9-10}
&  & \multicolumn{2}{c}{\ttfamily cwm}  &  & \multicolumn{2}{c}{\ttfamily norm}  &  & \multicolumn{2}{c}{\ttfamily pmm}\\
\cline{3-4} \cline{6-7} \cline{9-10}
&  & $\text{KL}_{\text{int}}$ & Relative distance &  & $\text{KL}_{\text{int}}$  & Relative distance &  & $\text{KL}_{\text{int}}$ & Relative distance\\ 
\cline{1-1} \cline{3-4} \cline{6-7} \cline{9-10} 
Qu.int. 95.0\% &  & (0,0.0055) & - &  & (0,0.0055) & - &  & (0,0.0055) & - \\ 
\cline{1-1} \cline{3-4} \cline{6-7} \cline{9-10} 
$\phantom{-}\bm{g}_{\bm{Y}_{\text{com}}}$  &  & 0.0029 & \phantom{-}{\ttfamily WI}  &  & 0.0029 & \phantom{-}{\ttfamily WI}  &  & 0.0029  & \phantom{-}{\ttfamily WI}   \\
$\phantom{-}\bm{g}_{\bm{Y}_{\text{obs}}}$  &  & 0.0679  & 12.25  &  & -  & -  &  & -  & - \\
$\phantom{-}\bm{g}_{\bm{Y}_{\text{mean}}}$ &  & 0.0854  & 15.42  &  & -  & -  &  & -  & - \\
$\phantom{-}\bm{g}_{\bm{Y}_{{X_1}}}$  &  & 0.0665  & 12.00  &  & 0.2093  & 37.79  &  & 0.1130  & 20.39 \\
$\phantom{-}\bm{g}_{\bm{Y}_{{X_2}}}$  &  & 0.0034  & \phantom{-}{\ttfamily WI}  &  & 0.0193     & \phantom{-}3.48  &  & 0.0022 & \phantom{-}{\ttfamily WI} \\
$\phantom{-}\bm{g}_{\bm{Y}_{{(X_1,X_2)}}}$  &  & 0.0036 & \phantom{-}{\ttfamily WI} &  & 0.0128  & \phantom{-}2.31  &  & 0.0040  & \phantom{-}{\ttfamily WI} \\ 
\cline{1-1} \cline{3-4} \cline{6-7} \cline{9-10} 
\end{tabular}%
}%

\caption{Performance of the {\ttfamily mean}, {\ttfamily cwm}, {\ttfamily pmm} and {\ttfamily norm} methods by calculating the KL divergence for the first simulated data set.}
\label{tabla:cwm_norm_pmm_simulacion1}
\end{table}

\subsubsection{A scenario with better performance of {\ttfamily cwm} versus {\ttfamily pmm}}
To further investigate the differences between the methods, a new database containing $n = 1000$ observations of the form $(x,y)$ was generated.
The mixing probabilities are $\alpha_1=0.6$ and $\alpha_2=0.4$, the mean vectors $\bm{\mu}_1=(4{.}0,10{.}0)$ and $\bm{\mu}_2=(7{.}0,4{.}0)$, and the covariance matrices are: 
\begin{equation*}
	\Sigma_1=\begin{pmatrix} 
	\phantom{-}0{.}50 & \phantom{-}0{.}35\\ 
	\phantom{-}0{.}35 & \phantom{-}0{.}50   
	\end{pmatrix} \quad \text{and} \quad \Sigma_2=\begin{pmatrix} 
	\phantom{-}0{.}50 & -0{.}64 \\ 
	-0{.}64 & \phantom{-}1{.}00  
	\end{pmatrix}.
\end{equation*}
The missing data pattern was generated in such a way that all observations with values of the variable $X$ greater than $5$ belonging to cluster 2 were considered as missing.  A summary of how the data was generated is presented in Table~\ref{cuadro:datos_simulados2}.
\begin{table}[htb]
\centering
\resizebox{11.5cm}{!}{%
\begin{tabular}{c|cl|cl|cl}
\multicolumn{1}{l|}{}   & \multicolumn{2}{c|}{\textbf{observed}}   & \multicolumn{2}{c|}{\textbf{missing}} & \multicolumn{2}{c}{\textbf{complete}} \\ \hline		\multirow{2}{*}{\textbf{cluster 1}}  & 441   & \textit{(76.7\%)} & 134 & \textit{(23.3\%)} & 575 & \textit{(100\%)} \\
& \multicolumn{1}{l}{\textit{(50.9\%)}} &  & \multicolumn{1}{l}{\textit{(100.0\%)}} &  & \multicolumn{1}{l}{\textit{(57.5\%)}} & \\ 
\hline
\multirow{2}{*}{\textbf{cluster 2}} & 425    & \textit{(100.0\%)} & \phantom{-}0  & \textit{(0.0\%)} & 425 & \textit{(100\%)} \\
& \multicolumn{1}{l}{\textit{(49.1\%)}} &   & \multicolumn{1}{l}{\textit{(0.0\%)}} &    & \multicolumn{1}{l}{\textit{(42.5\%)}} 
& \\ 
\hline
\multirow{2}{*}{\textbf{total}}  & 866  & \textit{(86.3\%)} & 134 & \textit{(13.4\%)} & 1000  & \textit{(100\%)} \\
& \multicolumn{1}{l}{\textit{(100\%)}}  &   & \multicolumn{1}{l}{\textit{(100\%)}}  &   & \multicolumn{1}{l}{\textit{(100\%)}}  &       
\end{tabular}
}%

\caption{Distribution of observed, missing and complete data by cluster. A scenario with censored data.}
\label{cuadro:datos_simulados2}
\end{table}

\begin{figure}[htb]
	\includegraphics[width=1.00\textwidth]{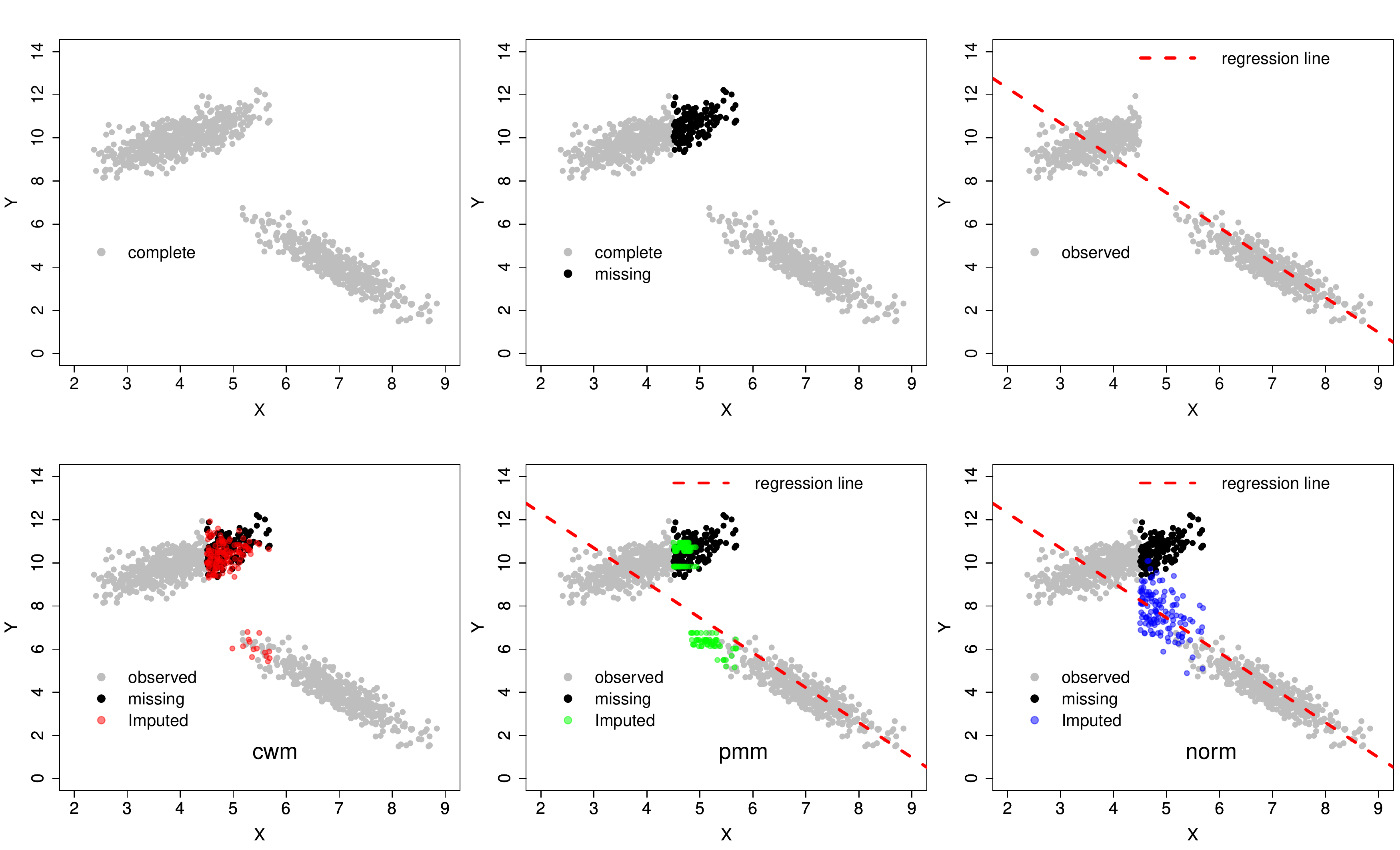}
	\centering
	\caption{Scatter plots for the {\ttfamily mean}, {\ttfamily cwm}, {\ttfamily pmm}, and {\ttfamily norm} methods in the case of censored missing data.}
	\label{scater_cwm_pmm_norm}
\end{figure}

The imputation process was implemented using the four methodologies of interest, {\ttfamily cwm}, {\ttfamily norm}, {\bf \ttfamily mean} and {\ttfamily pmm}.  Scatter plots for observed, missing, and imputed data by the four methods are shown in Figure \ref{scater_cwm_pmm_norm}. The panel in the upper left of Figure \ref{scater_cwm_pmm_norm} shows the results of the imputation process with the {\ttfamily mean} method. Since the methodology does not use information from the variable X, the procedure imputes in each cluster proportional to the number of data observed in each case, and imputes around the mean value of each group. For the {\ttfamily cwm} imputation procedure, we can see that the imputed values manage to cover a large part of the region where the missing data was generated, except for some imputations that occur in cluster 2. Regarding the {\ttfamily pmm} method, although some imputed values appear in the region of the missing data, they do not cover the region properly and a considerable number of aligned points are observed, which means that the method imputes with the same observed value many times. This situation occurs when there is little or no information observed in the specific region of imputation \citep{van2018flexible}. Additionally, a considerable amount of points were incorrectly imputed in cluster 2. The panel at the bottom right hand side shows the results of the imputation process with the {\ttfamily norm} method. The procedure erroneously imputes the vast majority of observations, specifically it does so in a region where there is no missingness.\\

From the analysis presented, it can be concluded that the best performance in terms of the imputation process was the {\ttfamily cwm} methodology. The {\ttfamily mean}, {\ttfamily pmm}, and {\ttfamily norm} methods perform unfavorably in this scenario. A quantitative evaluation of the imputation processes can be carried out from the KL divergence. Table \ref{Cuadro:KL-cwm-pmm-norm} presents this measure for the different methods to be compared. The 95\% quantile interval for the KL divergence of databases generated with the described specifications is shown in the first line. We assume that a KL divergence value that is within the interval allows us to conclude that the original distribution has been recovered, this will be noted in the relative distance column with {\ttfamily WI} (\textit{within the interval}).\\

The values in Table \ref{Cuadro:KL-cwm-pmm-norm} confirm the visual analysis previously performed. We see that the imputation made with {\ttfamily cwm} allows us to recover a sample from the original distribution, leading to the best performance among the other methods. The {\ttfamily norm} procedure shows the worst performance among the compared methods.\\
\begin{table}[htb]
\centering	
\begin{tabular}{llcclcc}
\cline{3-4}
\multicolumn{1}{l}{} &  & \multicolumn{2}{c}{\textbf{Approach method}}\\ 
\cline{3-4} 
\multicolumn{1}{l}{} &  & \multicolumn{1}{l}{$\text{KL}_{\text{int}}$} & \multicolumn{1}{l}{Relative distance}\\ 
\cline{1-1} \cline{3-4}
Qu.int. 95.0\% &  & (0, 0.0271)   & -  \\ 
\cline{1-1} \cline{3-4} 
\phantom{-----}$\bm{g}_{\bm{Y}_{\text{com}}}$  &  & 0.0028 & \phantom{-}0.10 \\
\phantom{-----}$\bm{g}_{\bm{Y}_{\text{obs}}}$  &  & 0.0611 & \phantom{-}2.25 \\
\phantom{-----}$\bm{g}_{\bm{Y}_\text{cmw}}$    &  & 0.0198 & \phantom{-}{\ttfamily WI} \\
\phantom{-----}$\bm{g}_{\bm{Y}_\text{pmm}}$    &  & 0.0559 & \phantom{-}2.06 \\
\phantom{-----}$\bm{g}_{\bm{Y}_\text{norm}}$   &  & 0.2174 & \phantom{-}8.02 \\
\cline{1-1} \cline{3-4} 
\end{tabular}	
\caption{KL divergences and relative distances for the imputation methods in the case of censored missing data.}
\label{Cuadro:KL-cwm-pmm-norm}	
\end{table}

An extreme scenario is presented where the pattern of missing data was generated considering censored data and where the distribution of the variables is separated by groups. The two clusters were generated from opposite correlations, one group with positive correlation and the other negative. In this situation, the {\ttfamily norm} and {\ttfamily pmm} imputation methods were not able to provide an adequate performance in imputation while the {\ttfamily cwm} was. Therefore, and based on all of our simulations studies, we conclude the {\ttfamily cwm} method is a interesting tool to perform imputation when inputs are available. First it was shown that, using appropriate inputs, it was able to provide proper sampling under the MAR mechanism. Also, even in a MNAR mechanism, it was able to perform reasonably imputing data in the correct component. Finally, comparing with existing methods, it was show that the {\ttfamily cwm} is very competitive having a much better performance in the last scenario discussed.\\ 

\section{Illustrative example: Faithful dataset}
\label{sec:ejemplo}

The Faithful database is a classic data set on geyser eruptions \citep{hardle1991smoothing}. The data contains the waiting times between eruptions and the durations of the Old Faithful geyser eruptions in Yellowstone National Park, Wyoming, United States. In the database, each row represents an observed eruption of the Old Faithful Geyser. The {\ttfamily Faithful} data set is found in the \texttt{datasets} \textsf{R} package and consists of $272$ observations on $2$ variables, {\ttfamily eruptions} (represents the duration of the eruption in minutes) and {\ttfamily waiting} (represents the wait time in minutes until the next eruption).\\

We will use this database to exemplify the imputation capacity and robustness of our method and compare it with the other methods studied in Subsection~\ref{subsec:simulacion2} in a MNAR scheme. Many authors have studied this data and concluded that a mixture of $G=2$ components is adequate to model it \citep[e.g.,][]{benaglia2009mixtools, Cabral2012, prates2013mixsmsn}. Therefore, we 
considered here a Gaussian mixture model with $G=2$ components as the ``true'' model.\\

\begin{figure}[htb]
    \centering
	\includegraphics[width=0.8\textwidth]{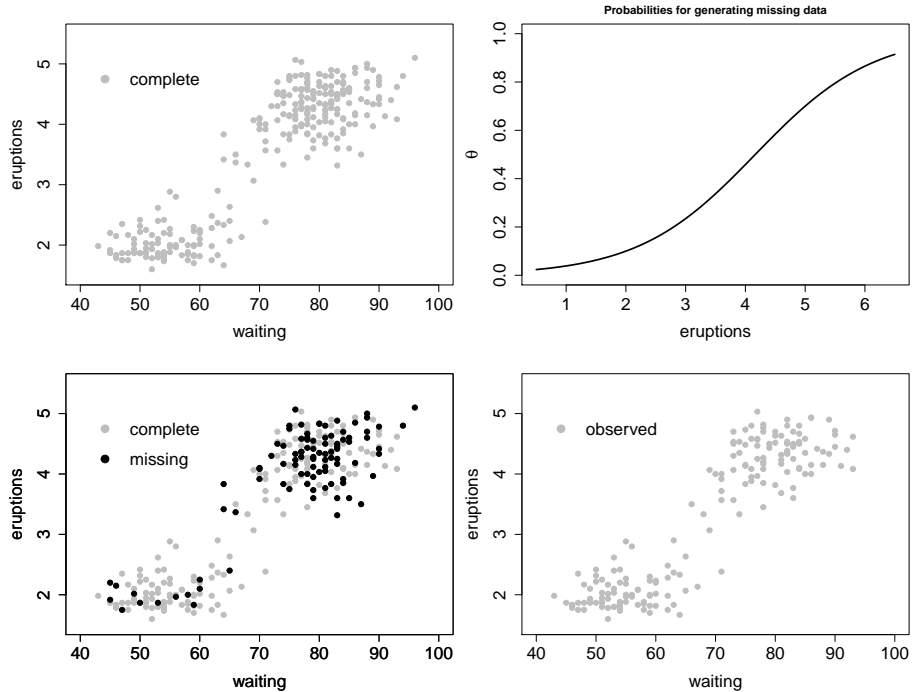}
	\caption{Construction of the missing data pattern for the {\ttfamily Faithful} dataset.}
	\label{fig:patern_mis_faithful}
\end{figure}

To generate the missing data, we assume a MNAR mechanism, that is, an indicator variable $R_i \sim \text{Bern}(\theta_i)$ is simulated with probability $\theta_i$ for the $i$-th observation depending on the output variable. This missing probability is obtained from the expression
\begin{equation}  
\theta_i=\text{logit}^{-1}(\beta_0+\beta_1 y_i) \qquad \mbox{for } i=1,...,n.
\label{logit}
\end{equation}
The value of $\theta_i$ is specified in such a way that the highest values of the output variable $y_i$ have higher probabilities of not being observed. Values of $\beta_0=-4.23$ and $\beta_1=1.02$ were used allowing larger values of the {\ttfamily eruptions} variable to have a greater probability of being missing. This scheme generated $95$ missing data out of the $272$ that the database has. Figure \ref{fig:patern_mis_faithful} presents the missing data structure with a scatter plot, and it also presents the probability graph to generate the missing data.\\

\begin{figure}[htb]
    \centering
	\includegraphics[width=1.0\textwidth]{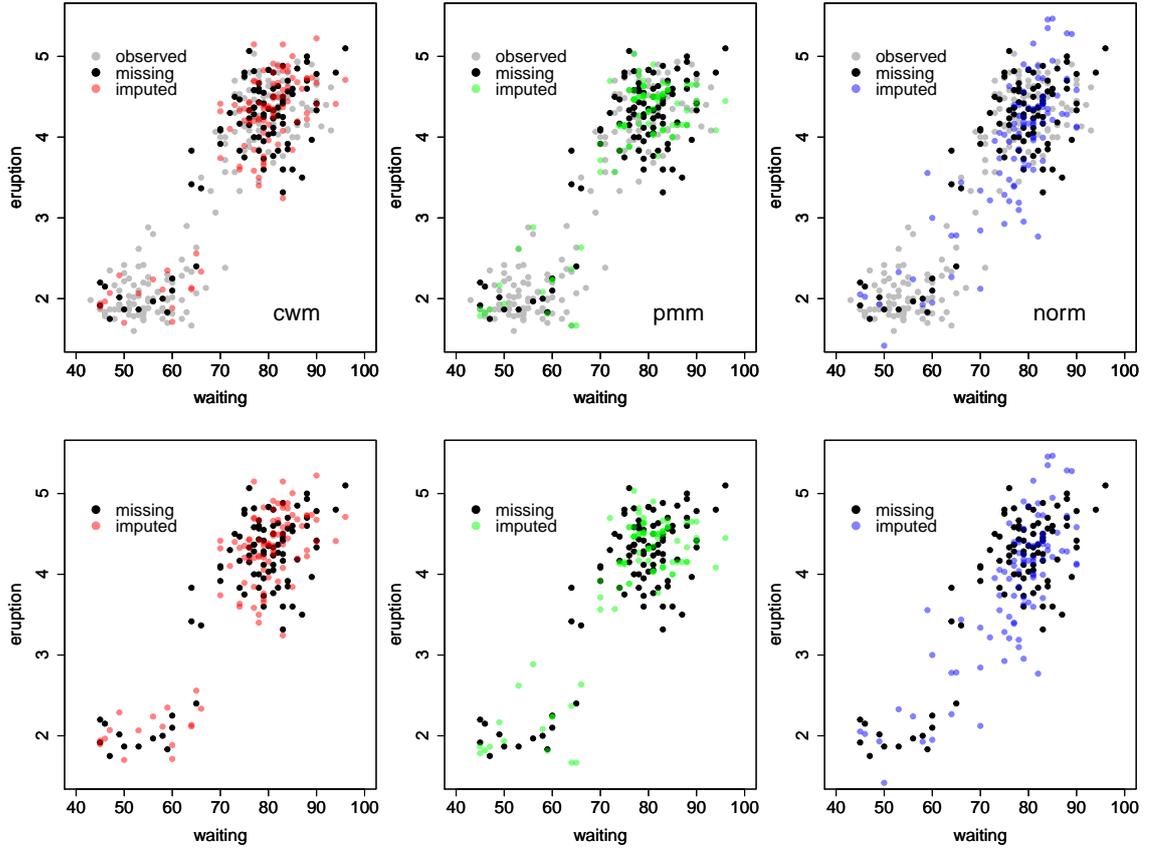}
	\caption{{\ttfamily Fainthful} dataset imputed using {\ttfamily mean}, {\ttfamily cwm}, {\ttfamily pmm}, and {\ttfamily norm} methods.}
	\label{fig:imp_cwm_pmm_norm_faithful}
\end{figure}

We proceed to impute the {\ttfamily Faithful} database, specifically the {\ttfamily eruptions} variable using the information from the fully observed {\ttfamily waiting} variable. Our model is compared to the {\ttfamily mean}, {\ttfamily pmm}, and {\ttfamily norm} procedures. Figure \ref{fig:imp_cwm_pmm_norm_faithful} presents scatter plots with the observed, missing, and imputed data for the four procedures. We can see that our model better covers the region of missing data. Some observations imputed with {\ttfamily pmm} are far from the missing data, while a considerable amount of data imputed with {\ttfamily norm} is imputed in the middle of the two components, in a region where there is no missing data nor observed. Graphically, we can see that of the four procedures presented, the one corresponding to {\ttfamily cwm} shows better performance, closely followed by {\ttfamily pmm}. Although visually the scatter diagram for the {\ttfamily mean} method presents the worst behavior, we must remember that this procedure does not use the information from the variable {\ttfamily waiting}, however it recognizes the existence of the two components and imputes, although in wrong proportion, in each one of them.\\


The graphical analysis carried out previously can be quantitatively corroborated using the KL divergence. Table \ref{tabla:KL_cwm-pmm-norm_faithful} calculates the divergence taking as reference the distribution of complete data. The relative distance is taken based on the KL divergence for the variable {\ttfamily eruptions} with only observed data. It is possible to conclude that the estimated distribution of the variable that loses the least information with respect to the complete data is the one imputed with our Gaussian LCWM proposal.

\begin{table}[htb]
\centering
\begin{tabular}{llcc}
\cline{3-4}
\multicolumn{1}{l}{}  &  & \multicolumn{2}{c}{\textbf{Approach method}}\\ 
\cline{3-4} 
\multicolumn{1}{l}{}  &  & \multicolumn{1}{l}{$\text{KL}_{\text{int}}$} & \multicolumn{1}{l}{Relative distance}  \\ 
\cline{1-1} \cline{3-4}  
$\bm{\text{{\ttfamily eruptions}}}_{\text{com}}$ &  & - & -  \\
$\bm{\text{{\ttfamily eruptions}}}_{\text{obs}}$ &  & 0.0297 & 1.00 \\
$\bm{\text{{\ttfamily eruptions}}}_{\text{mean}}$ &  & 0.0429 & 1.44 \\
$\bm{\text{{\ttfamily eruptions}}}_{\text{cwm}}$ &  & 0.0018 & 0.06  \\
$\bm{\text{{\ttfamily eruptions}}}_{\text{pmm}}$ &  & 0.0070 & 0.24  \\
$\bm{\text{{\ttfamily eruptions}}}_{\text{norm}}$ &  & 0.0518 & 1.74 \\
\cline{1-1} \cline{3-4}  
\end{tabular}
\caption{KL divergences in relation to the complete data distribution and its relative distances for the {\ttfamily Fainthful} dataset.}
\label{tabla:KL_cwm-pmm-norm_faithful}
\end{table}

\section{Conclusions}
\label{sec:conclusiones}

An imputation procedure that uses 
finite mixture models, specifically the Gaussian Linear Cluster-Weighted Modeling, is presented in this paper. 
Auxiliary information is included in the model through variables that are complete 
for all individuals. Under the assumption of starting from observational data, input and output variables are modeled jointly. This approach allows to calculate posterior probabilities that an individual belongs to each component given the auxiliary information. Through these probabilities, the model selects one of the various components and uses its conditional distribution to obtain imputations for the output variable using the input variables adaptively.\\

In the simulation study, we analyzed two extreme cases regarding the information provided by the input variable.
In the first scenario, the input variable does not provide any information on which component to impute from, and 
the imputation is determined by the mixing probabilities $\alpha_Z$ that are similar to the ones in the observed data. 
The second scenario occurs when the input variable is capable of selecting 
which component to impute from. This situation occurs, intuitively speaking, when the input variable 
separates the data distribution between components. In this last scenario, the proposed model has an optimal behavior when imputing the data. Any variable included in the model as auxiliary information will move between this range. 
This first simulation allowed us to verify that, when the auxiliary variable is capable of providing information about the components, the imputation was adequate. However, if the auxiliary variable does not have information about the components, the method cannot learn from it and the imputation will mimic the observed data (like the {\ttfamily mean} method). Finally, it was observed that when an input vector is used with one variable carrying useful information and another one not, the model is still able to impute adequately. In other words, the noise from the extra unnecessary variable does not affect the imputation capability of the proposed method.\\

A scenario where the missing data mechanism is MNAR was further simulated. Specifically, missing data were simulated in a particular region of one of the components. 
The objective was to observe the capacity of imputation of the proposed model in a scenario that was not constructed for, in other words, is the method somewhat robust to a more challenging scheme? 
It was possible to conclude that the method went halfway with the information of the input variables. The imputed values were located at the right component, 
however, they covered the entire region and not specifically the part of such component where the missing data was generated.\\ 

Further, our imputation model was compared with other methods in the literature.
As shown in this simulation, our imputation procedure was not worse and in specific cases presented better results in comparison to the other methods. This reinforces the conclusion that our imputation alternative is a robust one and should be considered when imputing data with auxiliary information.\\ 

For our application example, the Faithful database was used.
A missing pattern was simulated using an MNAR mechanism. Again, we were able to observe that, although our model imputes under the MAR assumption, including auxiliary information allowed it to have a good performance with respect to this more complex type of missingness mechanism, specially in comparison to the other methods. Thus, we conclude that our procedure is a strong candidate to perform imputation in comparison to the various mechanisms. The capacity of approximating the complete real data after imputing from the {\ttfamily cwm} was notable compared to the {\ttfamily mean} and {\ttfamily norm}. There was also improvement in comparison of the {\ttfamily pmm} procedure but it was less marked.\\

A direct future work of this study is to extend the methodology to the multivariate output variable case. Although no restrictions were made for the dimension of the input variable,
the idea is to extend the imputation methodology for the case of a $p-$dimensional output vector.\\ 
It would also be of interest to develop criteria for the selection of meaningful variables that can be used as auxiliaries, and that allows making the best possible imputation. Another possible extension is to create a {\ttfamily cwm-pmm} hybrid that takes advantage of both methodologies and check if it would improve the imputation performance. Finally, in some situations, according to the researcher's knowledge, it might be of interest to allow imputation to be carried in a particular region of the data set. Therefore, an extension to properly having a model that allows for explicit MNAR imputation is of interest.

\subsection*{Acknowledgement}

This study was financed in part by the Coordenação de Aperfeiçoamento de Pessoal de Nível Superior - Brasil (CAPES). Prates acknowledges partial funding support from Conselho Nacional de Desenvolvimento Científico e Tecnológico (CNPq) grants 436948/2018-4 and 307547/2018-4. Masmela-Caita acknowledges partial funding support from Universidad Distrital Francisco José de Caldas.

 \newpage

\bibliography{bibliografia.bib}   

 \newpage

\appendix

\section{Univariate simulation with information from an input vector $ (X_1, X_2)$}
\label{apendice_A}

\vspace{2cm}

In the simulation studies in Section 4.1, where we evaluate the performance of the variables that enter the imputation procedure under the Gaussian LCWM, we discuss the case of auxiliary information from an input vector. The vector is made up of the two variables that were considered in the analysis, that is, $\bm{X} = (X_1, X_2)$. Here are the graphs that illustrate this scenario.

\begin{figure}[htb]
\centering

\subcaptionbox{Construction of the model: $X_1 \times Y$ projection plane.
\label{subfig:base_1_uni_constX1X2_X1Y}}{\includegraphics[scale=0.250]{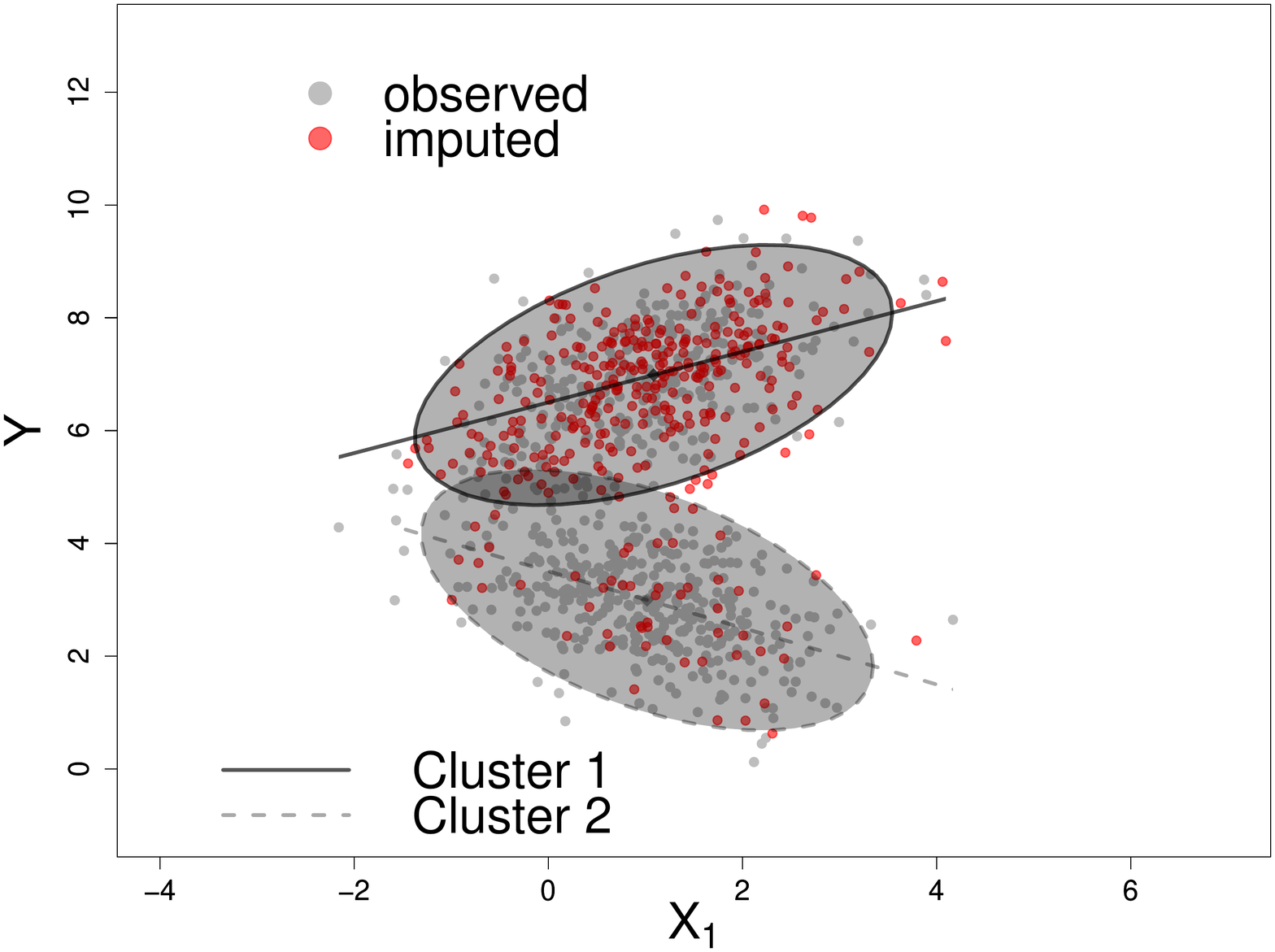}}
\subcaptionbox{Construction of the model: $X_2 \times Y$ projection plane.
\label{subfig:base_1_uni_constX1X2_X2Y}}{\includegraphics[scale=0.250]{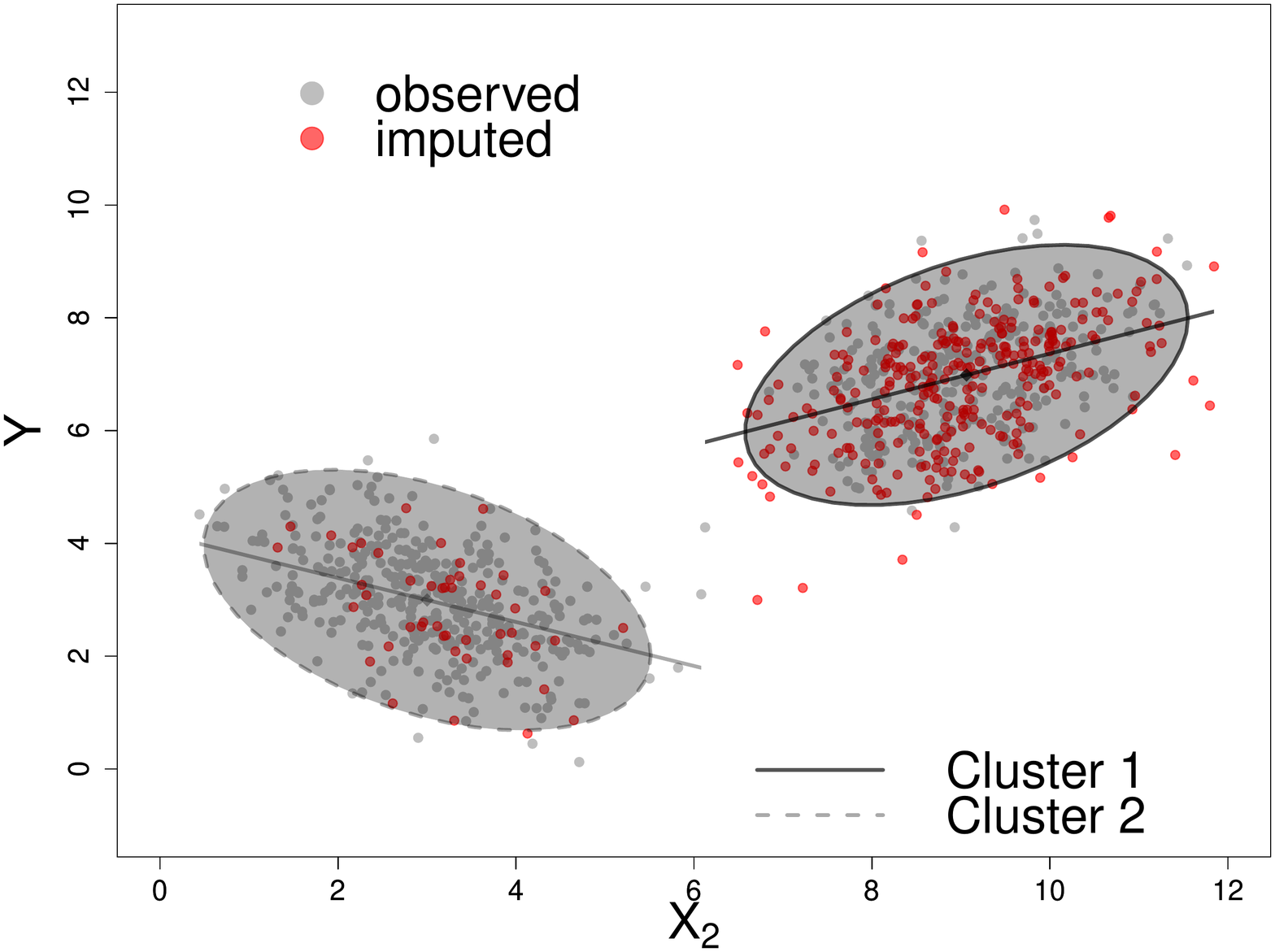}}
\subcaptionbox{Imputed data: $X_1 \times Y$ projection plane.
\label{subfig:base_1_uni_impX1X2_X1Y}}{\includegraphics[scale=0.250]{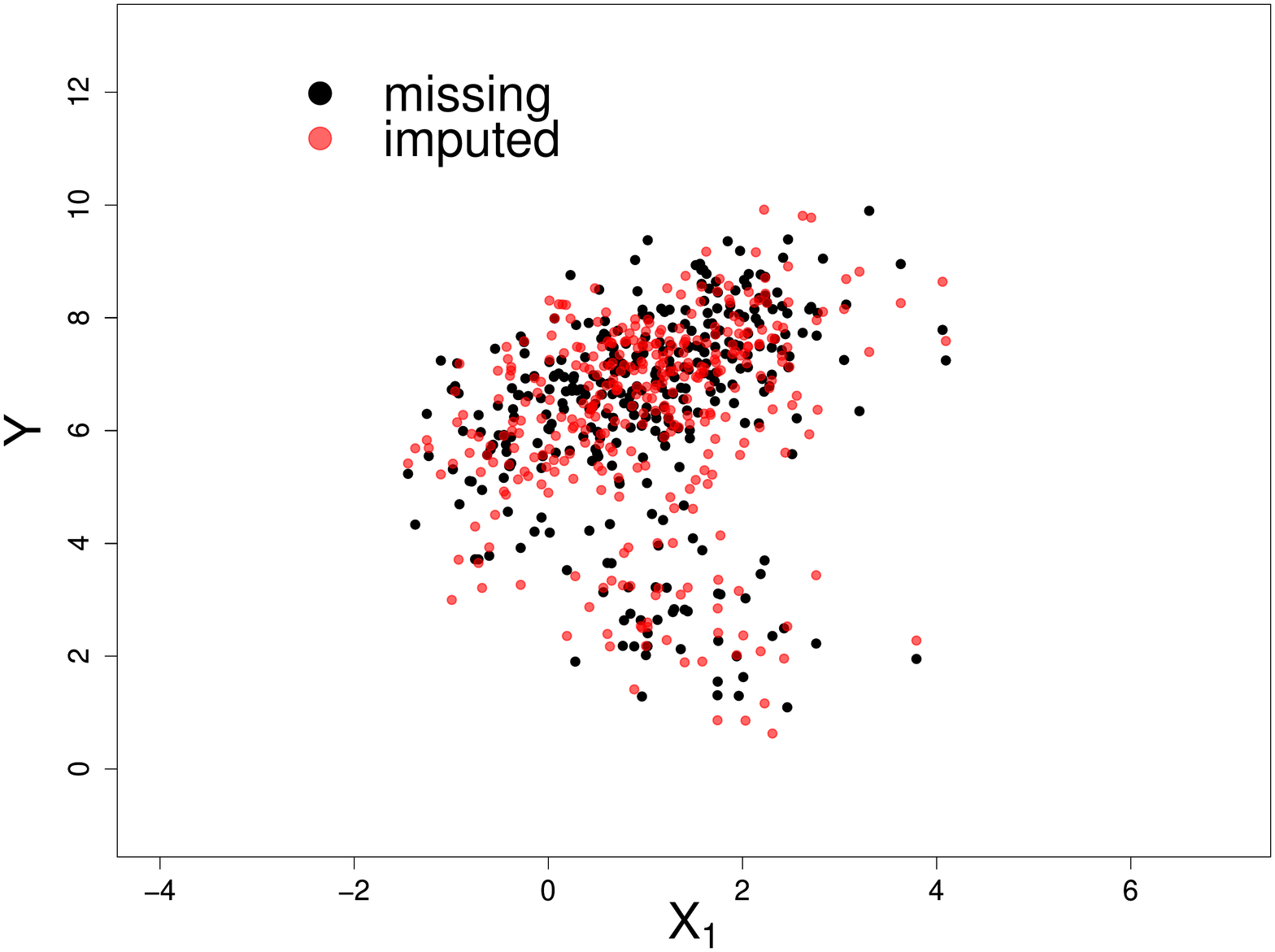}}
\subcaptionbox{Imputed data: $X_2 \times Y$ projection plane.
\label{subfig:base_1_uni_impX1X2_X2Y}}{\includegraphics[scale=0.250]{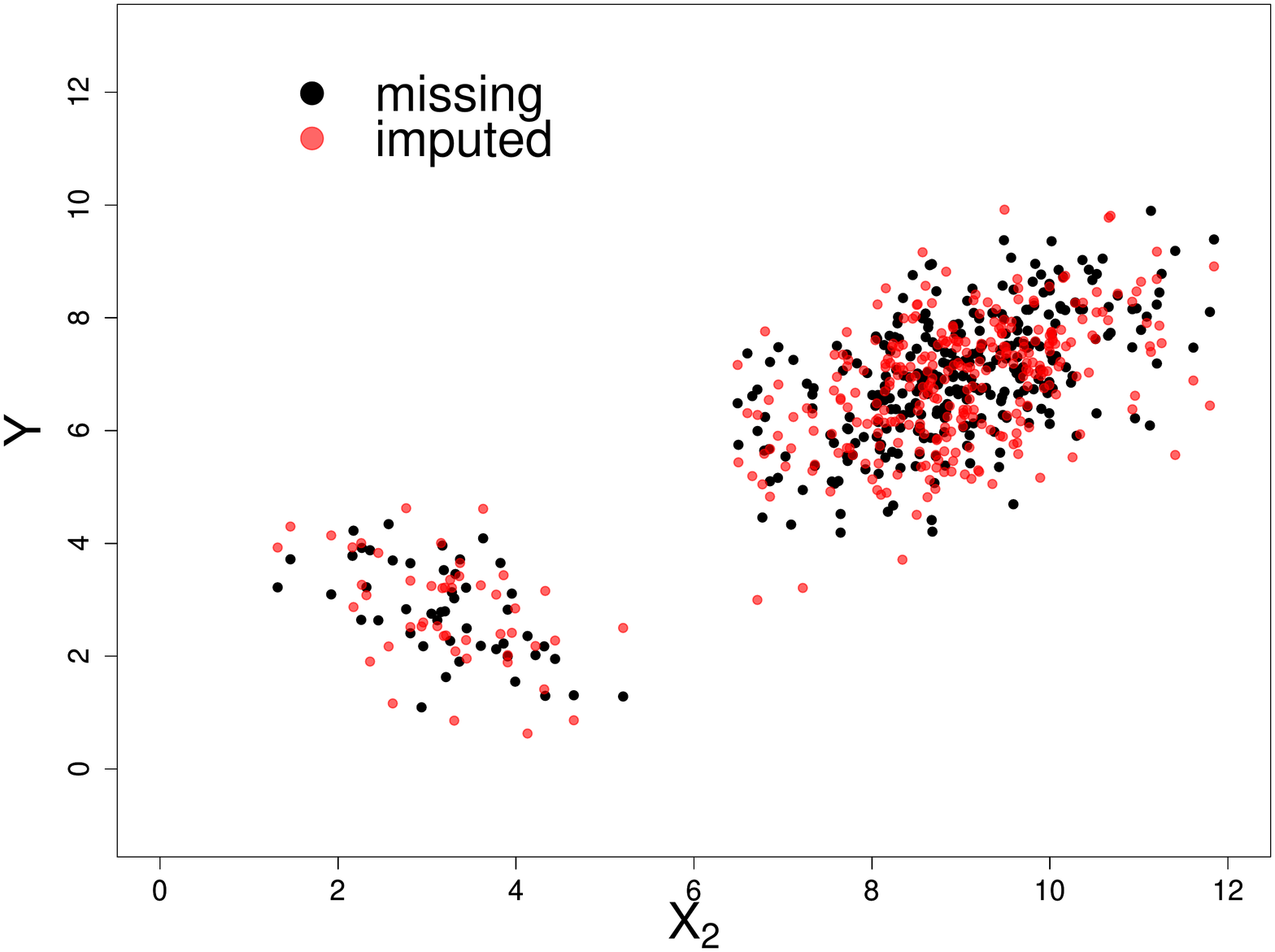}}
\caption{Construction of the univariate imputation model for simulated data considering the vector $(X_1, X_2)$ as auxiliary information.}
\label{fig:modelo_uni_X1_X2}
\end{figure}

\begin{figure}[htb]
\centering
\subcaptionbox{Surface $\alpha_1(x_1,x_2)$ corresponding to cluster 1.
\label{subfig:alpha_1}}{\includegraphics[scale=0.250]{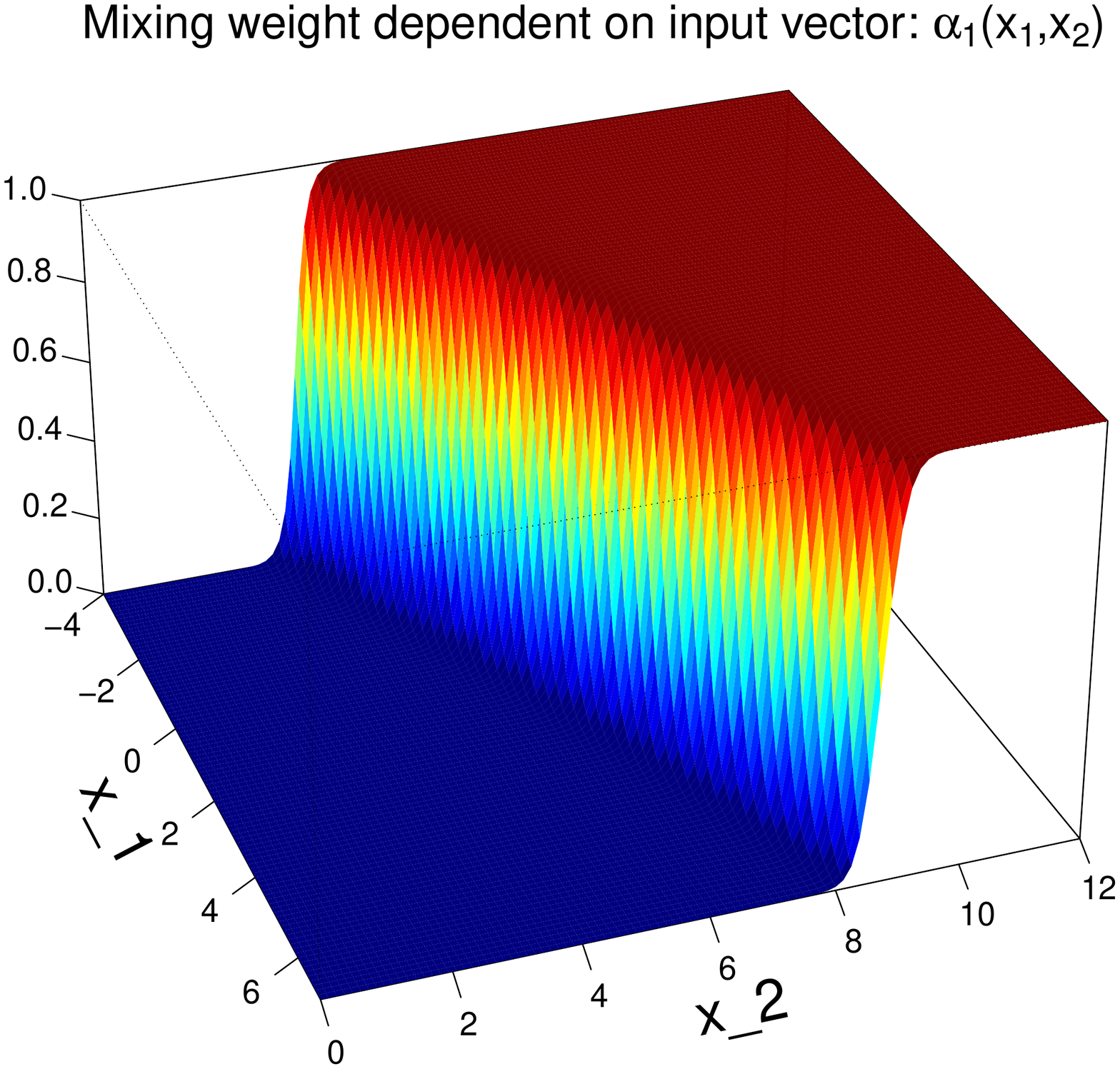}}
\subcaptionbox{Surface $\alpha_2(x_1,x_2)$ corresponding to cluster 2.
\label{subfig:alpha_2}}{\includegraphics[scale=0.250]{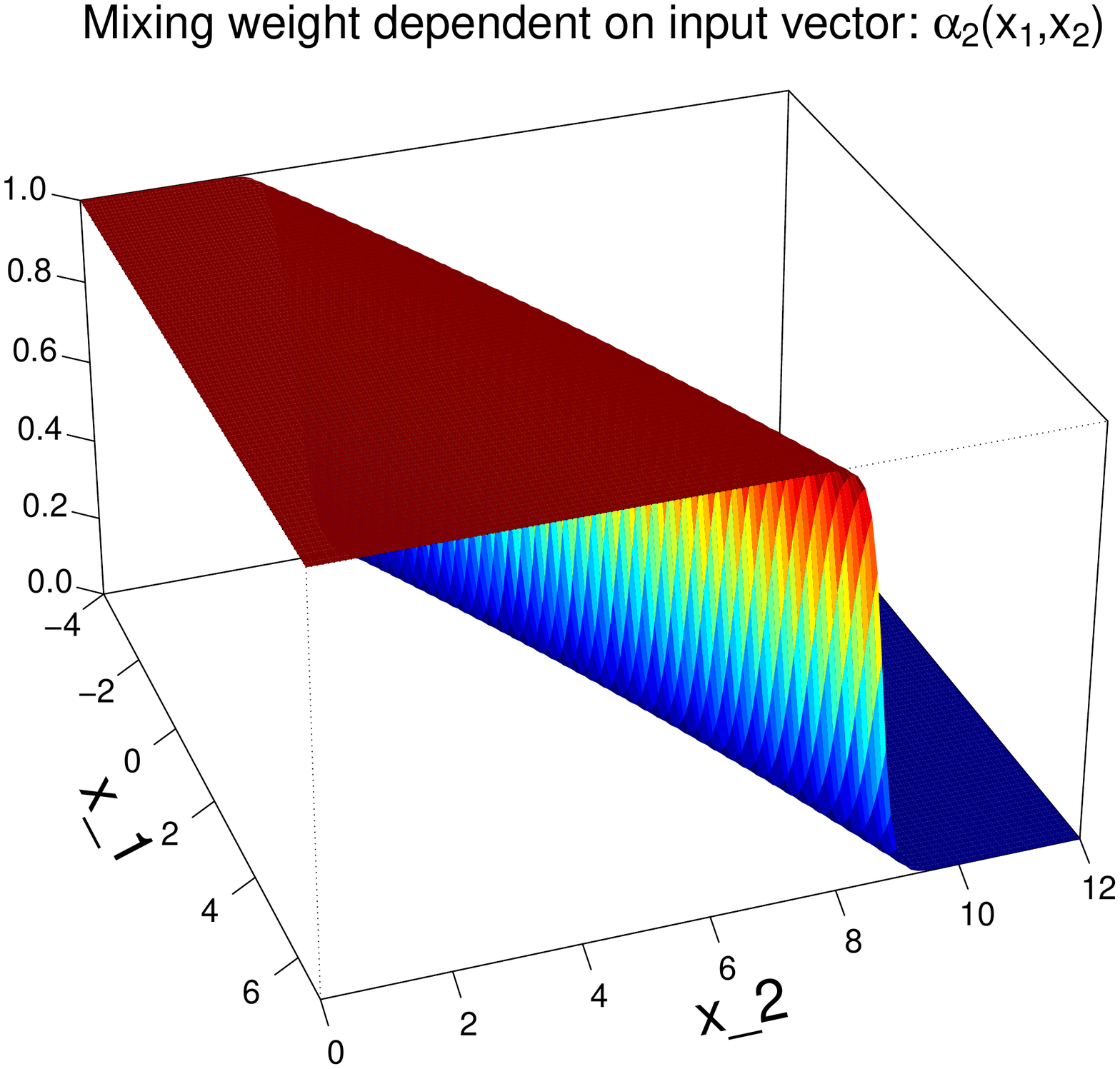}}
\subcaptionbox{Scatter plot for the complete data: $X_1 \times X_2$ plane projection
\label{subfig:proyeccion_X1X2}}{\includegraphics[scale=0.250]{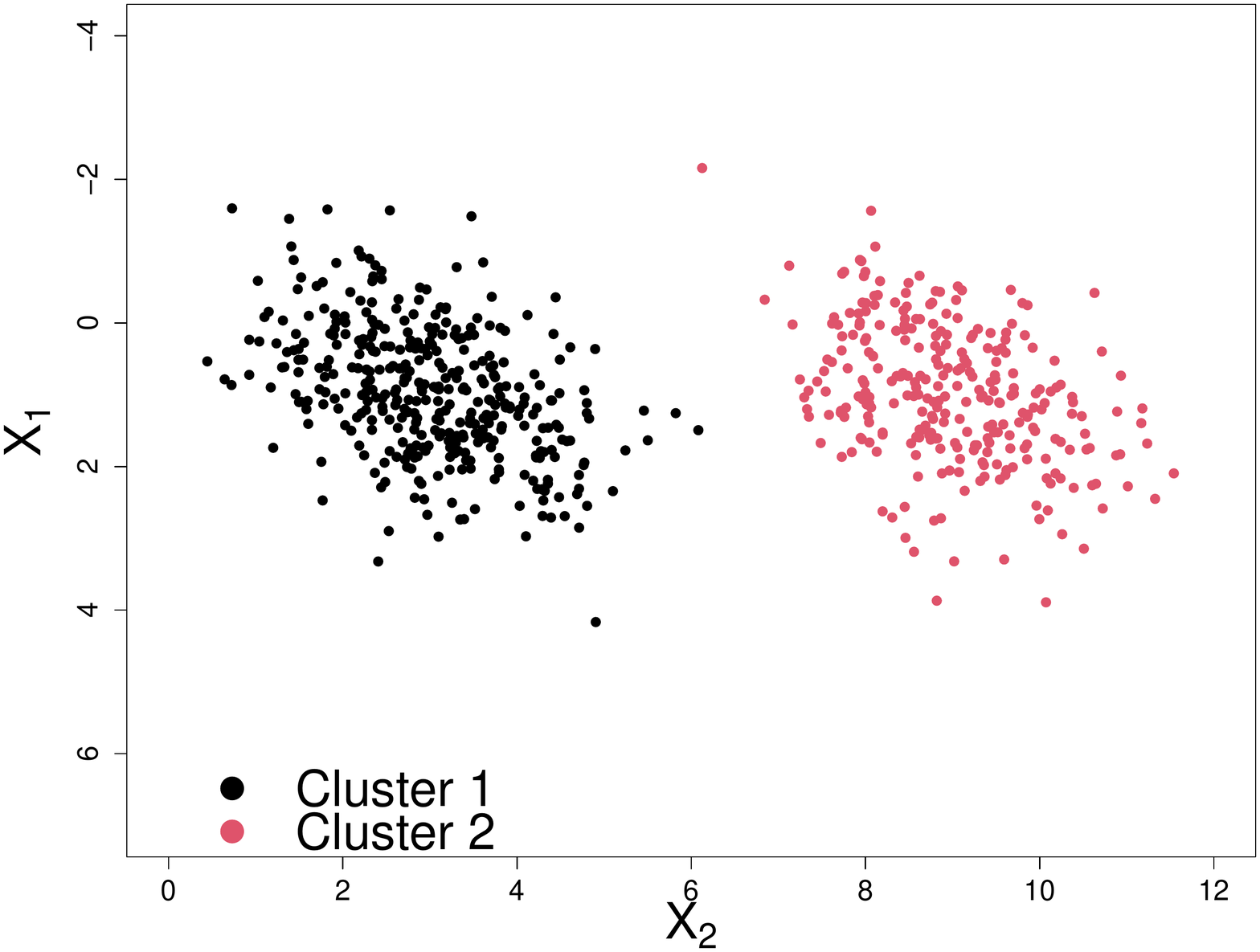}}

\caption{Mixture weights dependent on input vector $(X_1,X_2)$ for the construction of the univariate imputation model.}
\label{fig:prob_mezcla_X1_X2}
\end{figure}

\end{document}